\documentclass[12pt]{article}
\usepackage{amsmath, amsfonts, amssymb, amsthm, natbib}
\usepackage{graphicx,psfrag,epsf}
\usepackage{latexsym}
\usepackage{calc}
\usepackage[ragged, symbol]{footmisc}
\usepackage{grffile}
\usepackage{booktabs}
\usepackage{algorithmic}
\usepackage{algorithm}
\usepackage{caption}
\usepackage{comment}
\usepackage{color,url}
\usepackage{enumitem}
\usepackage{bbm}
\usepackage{adjustbox}

\usepackage{bm}

\newcommand{\V}[1]{{\bm{\mathbf{\MakeLowercase{#1}}}}} 
\newcommand{\M}[1]{{\bm{\mathbf{\MakeUppercase{#1}}}}} 
\newcommand{\T}{\intercal}
 
\newcommand{\blind}{1}

\usepackage{array}
\usepackage{makecell}

\theoremstyle{plain}
\newtheorem{remark}{Remark}[section]
\newtheorem{theorem}{Theorem}[section]
\newtheorem{lemma}{Lemma}[section]
\newtheorem{corollary}{Corollary}[section]

\begin{document}

\def\spacingset#1{\renewcommand{\baselinestretch}%
{#1}\small\normalsize} \spacingset{1}


\if1\blind
{
  \title{\bf Latent-state models for precision medicine}
  \author{Zekun Xu\\
  Department of Statistics, North Carolina State University\\
    and \\ 
    Eric B. Laber\thanks{
    The authors gratefully acknowledge funding from the National Science Foundation (DMS-1555141, DMS-1513579)
    and the National Institutes of Health (R01-DK-108073, P01 CA142538).
    }\hspace{.2cm}\\ 
    Department of Statistics, North Carolina State University\\
    and \\
    Ana-Maria Staicu \\
    Department of Statistics, North Carolina State University\\
    and \\
    Emanuel Severus \\
    Department of Psychiatry and Psychotherapy, Technische Universit\"at Dresden}
  \maketitle
} \fi

\if0\blind
{
  \bigskip
  \bigskip
  \bigskip
  \begin{center}
    {\LARGE\bf Latent-state models for precision medicine}
\end{center}
  \medskip
} \fi

\bigskip
\begin{abstract}
Observational longitudinal studies are a common means to evaluate 
  treatment efficacy and safety in chronic mental illness.  In many such
  studies, treatment changes may be initiated either by the patient receiving care or
  by their clinician and can thus vary widely across patients in 
  their timing,
  number, and type.  Indeed, in the observational longitudinal pathway
  of the STEP-BD study of bipolar depression, one of the motivations
  for this work, no two patients have the same treatment history even
  after coarsening clinic visits to a weekly time-scale.  Estimation
  of an optimal treatment regime using such data is challenging as one
  cannot naively pool together patients with the same treatment
  history as is required by methods based on inverse probability
  weighting or backwards induction. Thus, additional
  structure is needed to effectively 
  borrow information across patients and within a
  patient over time.  Current scientific theory for many chronic
  mental illnesses maintains that a patient's disease status can
  be conceptualized as transitioning among a small
  number of discrete states. We use this theory to 
  inform the construction of a partially observable Markov 
  decision process model of
  patient health trajectories wherein observed health
  outcomes are dictated by a patient's latent health state.
  Using this model, we derive and evaluate estimators of an optimal treatment regime under
  two common paradigms for quantifying long-term patient health.  
  The finite sample performance of the proposed estimator
  is demonstrated through a series of simulation experiments
  and application to the observational pathway of the STEP-BD
  study. We find that the proposed method provides high-quality
  estimates of an optimal treatment strategy in settings where
  existing approaches cannot be applied without {\em ad hoc}
  modifications.  
\end{abstract}

\noindent%
\vfill

\newpage
\spacingset{1.45} 

\section{Introduction}

A treatment regime is a set of decision rules that determines a personalized treatment plan  
by mapping a patient's evolving treatment and covariate history to a series of recommended treatments  
\citep[][]{murphy2003optimal, chakraborty2014dynamic,butchBookDTR}. 
An optimal
treatment regime maximizes the mean of some cumulative measure
of patient health when applied in a target population. Thus,
there is keen interest in the development of statistical
methodology for the estimation of optimal treatment regimes 
both to inform clinical practice and to generate new hypotheses
about heterogeneous treatment effects 
\citep[][]{athey2016recursive, wager2018estimation}.  
Seminal methods  for
estimating optimal treatment regimes from observational or 
randomized studies include g-estimation 
\citep[][]{robins1997, murphy2003optimal,robins2004optimal},
$Q$-learning and its variants 
\cite[][]{murphy2005generalization,moodie2007demystifying,henderson2010regret,schulte2014q,moodie2014q,song2015penalized,taylor2015reader,zhang2018interpretable,ertefaie2014constructing}, 
and inverse probability weighting 
 \citep[][]{robins1999testing, murphy2001marginal,van2006causal,robins2008estimation}.  More recently,
there has been a surge of research on extending these methods
to make them more flexible, e.g., through the use of
machine learning methods \cite[][]{zhao2012estimating,zhang2012estimating,rubin2012statistical,moodie2014q,zhao2009reinforcement,zhao2015new,laber2015tree,luedtke2016super,xu2016bayesian,wager2018estimation,tao2018tree,jiang2019entropy,luckett2019estimating,liu2019learning}, 
or to allow them to
work with high-dimensional feature spaces or other complex
data structures \cite[][]{lu2013variable,mckeague2014estimation,tian2014simple,song2015penalized,ciarleglio2015treatment,ciarleglio2016flexible,laber2017functional,shi2018high,ertefaie2014constructing,wallace2019model,shi2019sparse}.  While
the literature on treatment regimes is rich and growing rapidly,
the types of data to which current methods can be applied is restrictive.  
Existing  methods for finite-horizon decision problems 
require that one be able to align patient treatment decisions 
in time and that
the conditional average treatment effect at each decision point
be estimated using either  regression or weighting methods.  
For infinite- or indefinite-horizon problems, existing methods
require that the data-generating distribution
have sufficient structure to allow the pooling of data over time points and
extrapolation to future decisions, e.g., the data-generating model
might be assumed to be a contextual bandit or  
a homogeneous Markov decision process 
\citep[MDP,][]{tewari2017ads,ertefaie2014constructing,luckett2019estimating,liao2019off}.  
The setting we consider here has frequent and irregularly spaced
treatment changes so that patients cannot be aligned over time points 
nor is it clinically plausible that the data are Markov;  see Figure
\ref{stepBDTxtHistory}, which displays patient treatment
histories for a subset of patients from the STEP-BD standard care pathway. 

We propose a method for estimating an optimal treatment regime in the 
indefinite-horizon setting 
when data are irregularly spaced, contain multiple treatment
changes, and cannot be assumed to be Markov.  Motivated
by the underlying clinical science of bipolar depression and other 
episodic chronic illnesses, we assume that a patient's health
status is dictated by a latent (unobserved) state and a subset
of their observable data; we assume that conditional on current patient information
and this latent state, the evolution of a patient's health status is Markov.  Treatment
is allowed to affect the transition dynamics of the latent process as
well as patient observables.  We show that under this model the 
optimal treatment regime is determined by the so-called information
state, which comprises the conditional distribution of the latent
state and current patient measurements.  
We subsequently derive estimators of the optimal treatment regime and 
establish their asymptotic operating characteristics.  

The proposed model is an example of a partially observable
MDP \citep[POMDP,][]{monahan1982state}. POMDPs 
have been studied extensively in the computer science literature
with applications in robotics, scheduling, 
videogames, and wildlife management 
\cite[see, for example,][]{kaelbling1998planning,cassandra1998survey,pineau2003pbvi,hansen1998solving,ji2007point,sutton1998reinforcement}.  The primary contributions of this
work include: a theory-driven construction of the latent-process model,
the application of POMDPs to episodic chronic mental illness, 
and the development of valid statistical inference for clinically relevant estimands in this context.  The proposed methodology is extensible
and could be adapted for estimation and inference with optimal
treatment regimes in other contexts that have complex treatment 
and observation patterns,
e.g., mobile-health with habituation modeled as a latent process.  

The remainder of this manuscript is organized as follows.
In Section 2, we formally introduce the latent state 
model and show that the information state is, in some sense,  
minimally sufficient 
for the optimal treatment regime.   In Section 3, we review 
estimation of optimal treatment regimes under an MDP
model.  In Section 4, we derive estimators of the optimal 
treatment regime based on a data-driven transformation of the 
observed process which makes it approximately Markov and thus
amenable to the methods reviewed in Section 3. 
In Section 5, we derive the asymptotic distributions
of the proposed estimators.  In Section 6, we study the 
finite sample performance of the proposed methods through
an extensive suite of simulation experiments. In Section 
7, we present an illustrative application using the standard
care pathway of the STEP-BD bipolar disorder study.  We make
concluding remarks in Section 8.

\section{Setup and preliminary results}
We use uppercase letters, e.g., $\M{X},$ $T$, and $A$, to denote
random variables and lower case letters, e.g., $\V{x}$, $t$, and $a$, to
denote instances of these random variables.  
The symbol `$\triangleq$' is
used to distinguish definitions from equalities.  The observed data
are assumed to comprise $n$ i.i.d. copies, one per patient, of the
trajectory $\left\lbrace\left(T^j, A^j, \M{X}^j\right)
\right\rbrace_{j=1}^{J}$,
where $J\in\mathbb{Z}_+$ is the number of clinic visits, 
$0=T^1 < T^2 <\cdots < T^J \le 1$ encode clinic visit times;
$A^j = A(T^j)\in\mathcal{A}=\lbrace 1,\ldots, L\rbrace$ denotes the
assigned treatment during period $[T^j, T^{j+1})$; and 
$\M{X}^j = \M{X}(T^j) \in\mathbb{R}^p$ denotes a patient's health status at time
$T^j$, $j=1,\ldots, J$.  Thus, both the number and timing of 
clinic visits are treated as random quantities.  Let
$\M{H}^1 \triangleq \left\lbrace T^1, \M{X}^1\right\rbrace$ 
and $\M{H}^j \triangleq \left\lbrace \M{H}^{j-1}, A^{j-1}, T^j,  
\M{X}^j \right\rbrace$ so that $\M{H}^j$ contains the patient
history available to the decision maker at clinic visit $T^j$
before treatment $A^j$ is assigned.

Let $\mathcal{D}$ denote the space of probability distributions over $\mathcal{A}$
(i.e., the $L$-dimensional probability
simplex). We encode elements $\V{d}\in\mathcal{D}$ as vectors in
$[0,1]^L$ so that  $d_{a}$ represents the probability of selecting
treatment $a\in\mathcal{A}$ under $\V{d}$.   
A treatment regime in this setting is a sequence of decision rules
 $\M{\pi} = \left\lbrace \pi^j
\right\rbrace_{j\ge 1}$, one per clinic visit, with
$\pi^j:\mathrm{supp}\,\M{H}^j \rightarrow \mathcal{D}$, so that
under  $\M{\pi}$ a patient presenting with $\M{H}^j = \V{h}^j$ at
clinic visit $j$ would receive treatment recommendation $a$
with probability 
 $\pi_a^j(\V{h}^j)$.  Whereas the observed data comprise finite
 patient trajectories, we are interested in estimating treatment
 regimes that can be applied indefinitely; that is, they can
 be used to provide treatment recommendations for as long as the
 patient is receiving care.  To this end, we consider treatment
 regimes composed of decision rules $\pi^j = \rho \circ f^j$,
 where $\V{f} = \left\lbrace f^j\right\rbrace_{j\ge 1}$ are 
 summary functions $f^j:\mathrm{supp}\,\M{H}^j\rightarrow\mathcal{S}
 \subseteq \mathbb{R}^q$, so that $\M{S}^j \triangleq f^j(\M{H}^j)$ is a summary
 of patient history $\M{H}^j$, and $\rho:\mathcal{S}\rightarrow 
 \mathcal{D}$ is a stationary decision rule acting on patient summaries. 
 Write $\V{\pi} = \rho\circ \V{f}$ to denote the composed
 regime $\pi^j = \rho \circ f^j$ for all $j\ge 1$.  
 We will show below that restricting attention to composed 
 regimes of this
 type incurs no loss of generality. Furthermore, because
 $\rho$ remains fixed in this representation, the regime can be 
vetted for clinical 
 validity by domain experts when the summary functions 
 provide `qualitatively similar' summaries of the patient 
 history (we show that the natural choice of summary
 function in our domain produces such summaries).

We assume there exists a fixed real-valued function $\mathcal{U}:\mathcal{S}\times\mathcal{A}\rightarrow\mathbb{R}$
so that 
the immediate utility associated with history $\M{H}^j$ and
treatment $A^j$ is $U^j \triangleq \mathcal{U}(\M{s}^j, A^j)\in \mathbb{R}$.
It is thus assumed that the immediate utility depends on the history only through its summary;
note that  the summary function can always be chosen to ensure that this holds. 
Let $\mathbb{E}^{\pmb{\pi}}$ denote expectation with respect to 
the probability distribution
induced by following the treatment recommendations given by 
$\M{\pi}$ (for a formal development using potential outcomes, see the Supplemental Material; see also \citet{butchBookDTR}).  
We consider the following two measures of cumulative
utility:\\ (i)  discounted mean utility 
\begin{equation*}
V_{\mathrm{dis}}(\V{\pi}) \triangleq 
\mathbb{E}^{\V{\pi}}\sum_{j\ge 1}
\left(
\gamma^{j-1}U^j
\right),
\end{equation*}
where $\gamma \in [0,1)$ is a discount factor, and\\ (ii) average utility 
\begin{equation*}
V_{\mathrm{ave}}(\V{\pi}) \triangleq \lim_{N\rightarrow \infty}
\mathbb{E}^{\V{\pi}}
\left(
\frac{1}{N}
\sum_{j=1}^{N}U^j
\right).
\end{equation*}
These two cumulative measures are used almost exclusively in 
indefinite decision problems \citep[][]{powell2007approximate,busoniu2017reinforcement,sutton1998reinforcement}, though
the proposed methods could be extended to hyberbolic discounting
or other notions of cumulative utility \cite[][]{fedus2019hyperbolic}.
We write $V(\V{\pi})$ without a subscript to  denote generically either 
of these cumulative utility measures. 
We say that $\V{\pi}^{\mathrm{opt}}$
is optimal with respect to $V$ if $V(\V{\pi}^{\mathrm{opt}}) \ge 
V(\V{\pi})$ for all $\V{\pi}$.  
Without imposing additional structure on the data-generating model, 
it is not possible in general to identify $\V{\pi}^{\mathrm{opt}}$ 
 from data collected over a finite time-horizon even as $n\rightarrow \infty$.  

 We assume that there exists a (latent) Markov process 
 $M(t) \in \left\lbrace  1,\ldots, K \right\rbrace$ 
 indexed by $t\in [0,1]$ that 
 represents a critical component of a patient's health status,
 e.g., in of model of bipolar depression this represents whether the patient
 is in a depressive, manic, hypomanic, mixed, or stable episode at time $t$. 
Furthermore, we assume that:
\begin{itemize}
\item[(A1)] $\M{X}^j \perp M(T^1),A^1,\M{X}^1,\ldots,M(T^{j-1}),A^{j-1}
\big| M(T^j), A^j, \M{X}^{j-1}$ for all $j\ge 2$.  
\end{itemize}
so that the process is conditionally 
Markov given the latent state, i.e., given a summary of a patient's (observable) history, $\mathbf{X}^j$,
their latent health state, and current treatment, the future is independent of the past.    
The following result shows that
the conditional distribution of the latent state  given the available
history is sufficient for the optimal regime.  
\begin{lemma}\label{pomdpToMDP}
Assume (A1) and for each $j\ge 1$ let $\M{B}^j \in [0,1]^K$ be such that 
$B_{\ell}^j = P\left\lbrace M(T^j) = \ell\big|\M{H}^j\right\rbrace$
for $\ell=1,\ldots, K$.
Define $f^j(\M{H}^j) \triangleq 
\left(\M{B}^j, \M{X}^j\right)$ and write
$\M{S}^j = f^j(\M{H}^j)$ with $\mathcal{S} = \mathrm{supp}\,\M{S}^j$.
If $U^j$depends on $(\M{H}^j, A^j)$ only through $(\M{S}^j, A^j)$\footnote{
  As noted previously, the utility is typically a function of observables, and thus $\M{X}^j$ can 
  always be defined so as to include $U^j$.  However, this assumption also allows for utility to be the posterior
  of some latent patient characteristic given the history and treatment. 
} then: 
\begin{itemize}
  \item[(i)] $\left\lbrace \left(\M{S}^j, A^j, U^j\right)\right
  \rbrace_{j\ge 1}$ is a homogeneous MDP, and  
  \item[(ii)] $\sup_{\rho:\mathcal{S}\rightarrow \mathcal{D}}V(\rho
  \circ\V{f}) = \sup_{\M{\pi}}V(\M{\pi})$.
\end{itemize}
\end{lemma}

\begin{remark}
The summary $\M{S}^j$ is minimally sufficient in that there exists
generative models in which any further reduction of the
history, e.g., learning a strategy that depends on 
$\M{R}^j = g^j(\M{S}^j)$ where $||Cov(\M{S}^j|\M{R}^j)|| > 0\, 
\mathrm{w.p.1.}$, 
leads to degradation in the value of the learned strategy.
See the Supplemental Material for a precise statement and example.  
\end{remark}

Lemma \ref{pomdpToMDP} establishes that we can characterize the optimal regime
in terms  of the MDP
 $\left\lbrace (\M{S}^j, A^j, U^j)\right\rbrace_{j\ge 1}$,
 which admits a stationary optimal
regime, $\pi^{\mathrm{opt}} = \arg\max_{\rho:\mathcal{S}\rightarrow \mathcal{D}}
V(\rho\circ\V{f})$.  Thus, the structure provided by the MDP reduces the 
problem of estimating an optimal treatment regime from a search over the space
of countable sequences of functions, each acting on a different domain, to a 
search for a single function mapping $\mathcal{S}$ into $\mathcal{D}$ 
\citep[this is why, hereafter, we reference regimes using the unbolded
$\pi$; see][for additional discussion of the structure induced by MDPs]{sutton1997significance, sutton1998reinforcement}. 
Were trajectories from this MDP observed,  an estimated
optimal regime could be obtained by solving estimating 
equations based on the Bellman optimality conditions;
we review these estimating equations in the next section.
As the states $\M{S}^j,\, j\ge 1$ are not fully observed, we
first construct estimators 
$\left\lbrace \widehat{\M{S}}_n^j\right\rbrace_{j=1}^J$ of
$\left\lbrace \M{S}^j\right\rbrace_{j=1}^J$, and then
plug them into the MDP estimating equations. 
Asymptotic results for estimators
of this type are provided in Section 5.

\section{Optimal treatment regimes in an MDP}\label{dtr}
Recall that our approach is to transform the observed data
so that it mimics data collected under the homogeneous 
MDP of Lemma \ref{pomdpToMDP}.  To illustrate how this transformed
data will be used, we briefly review
two established methods for estimating an optimal treatment regime
in an MDP. Our developments closely follow  
\citet{murphy2016batch},
\citet{ertefaie2014constructing}, and 
\citet{luckett2019estimating}.   For a more general treatment
of MDPs see \citet{sutton1998reinforcement} and \citet{wiering2012reinforcement}. 
For the purpose of describing these methods, 
assume that the observed data are
$\left\lbrace (\M{S}_i^j, A_i^j, U_i^j), j=1,\ldots, J_i\right\rbrace_{i=1}^{n}$, 
which consist of $n$ independent trajectories from a 
homogeneous MDP.  We assume that the states take values in Euclidean space so that $\M{S}^j \in \mathcal{S}\subset 
\mathbb{R}^q$, there
are a finite number of treatment options coded  so that $A^j 
\in \left\lbrace
1,\ldots, 
L\right\rbrace$, and the utilities $U^j\in\mathbb{R}$ are coded so
that higher values are better.  
We present estimating equations for the optimal treatment regime 
with both the discounted and average utility criteria. Technical conditions needed for unbiasedness
of these estimating equations and asymptotic normality of the resultant estimators 
applied to the transformed data,
$\left\lbrace (\widehat{\M{S}}_{n,i}^j, A_i^j, U_i^j), j=1,\ldots, J_i\right\rbrace_{i=1}^{n}$, 
are provided in Section 5.  

For any regime $\pi$ and state $\V{s}$,
define the discounted state-value function 
$\nu_{\mathrm{dis}}(\pi, \V{s}) \allowbreak  = \\ \allowbreak
\mathbb{E}^{\pi}\left(
\sum_{j \ge 0}\gamma^{j}U^{t+j}\big|\M{S}^t = \V{s} 
\right)$, which does not depend on $t$ because the MDP 
is assumed to be homogeneous \citep[][]{puterman2014markov}.  
For any distribution $R$ on $\mathcal{S}$, termed
a reference distribution, define
$V_{\mathrm{dis}}^{R}(\pi) = \int \nu_{\mathrm{dis}}(\pi,\V{s})dR(\V{s})$, then
$V_{\mathrm{dis}}(\pi) = V_{\mathrm{dis}}^{R_0}(\pi),$ where 
$R_0$ is the initial state distribution.  Because $R_0$ is 
unknown, one might take the empirical distribution of $\M{S}^1$ or 
some other reference distribution constructed from historical data 
\citep[see][]{luckett2019estimating}.
Define 
$\pi_{\mathrm{dis}}^{\mathrm{opt},R} = \arg\max_{\pi\in\Pi}V_{\mathrm{dis}}^R(\pi)$, where $\Pi$ denotes a class of regimes of interest. 
In the discounted utility case, it can be shown 
\citep[e.g.,][]{luckett2019estimating}
that the state-value function satisfies the following recursion
\begin{equation}\label{popnEstEqVLearning}
0 = 
\mathbb{E}
\left[
\frac{
  \pi_{A^j}(\M{S}^j)
}{
  P(A^j\big|\M{S}^j)
}
\left\lbrace
U^j + \gamma \nu_{\mathrm{dis}}({\pi}, \M{S}^{j+1})
- \nu_{\mathrm{dis}}({\pi},\M{S}^j)
\right\rbrace  \phi(\M{S}^j)
\right],
\end{equation}
for all $j$ and any $\phi:\mathcal{S}\rightarrow \mathbb{R}^d$, where
the ratio $\pi_{A^j}(\M{S}^j)/P(A^j|\M{S}^j)$ is an importance sampling
weight.    
Let $\mathcal{V} = \left\lbrace
\nu_{\mathrm{dis}}(\M{S}^j;\V{\alpha})\,:\,\V{\alpha} \in \Omega\subseteq \mathbb{R}^d
\right\rbrace$ be a parametric class of continuously differentiable
maps from $\mathcal{S}$ into $\mathbb{R}^d$; we have overloaded the notation 
$\nu_{\mathrm{dis}}$ to reflect that each regime $\pi$ will be associated with a corresponding
parameter vector $\V{\alpha}$.   For each ${\pi} \in \Pi$,
define $\V{\alpha}^*({\pi})$ to be the solution to 
(\ref{popnEstEqVLearning}) at ${\pi}$.  An estimator
$\widehat{\V{\alpha}}_{n}({\pi})$ of $\V{\alpha}^*({\pi})$ is given 
by the solution of the sample analogue of (\ref{popnEstEqVLearning})
with $\phi(\M{S}^j) = \nabla_{\V{\alpha}}\nu(\M{S}^j;\V{\alpha})$, i.e., the
solution to  
\begin{equation}\label{sampleEstEqVLearning}
0 = 
\mathbb{P}_n\sum_{j=1}^{J-1}
\left[
\frac{
  \pi_{A^j}(\M{S}^j)
}{
  P(A^j\big|\M{S}^j)
}
\left\lbrace
U^j + \gamma \nu_{\mathrm{dis}}(\M{S}^{j+1};\V{\alpha})
- \nu_{\mathrm{dis}}(\M{S}^j;\V{\alpha})
\right\rbrace  \nabla_{\V{\alpha}} \nu_{\mathrm{dis}}(\M{S}^j;\V{\alpha})
\right],
\end{equation}
where $\mathbb{P}_n$ denotes the empirical measure.  The estimated optimal
regime is obtained by maximizing the estimated integrated state-value function 
over the class of regimes so that
\begin{equation*}
  \widehat{\pi}_{\mathrm{dis},n}^R = 
  \mathrm{argmax}_{\pi\in\Pi}\int \nu_{\mathrm{dis}}\left\lbrace 
  \V{s}; \widehat{\V{\alpha}}_n({\pi})\right\rbrace dR(\V{s}).
\end{equation*}
Properties of this estimator---applied to data from a homogeneous MDP---are provided
 in \citet{luckett2019estimating}.
We assumed for simplicity that $P(A^j|\M{S}^j)$ was known, e.g., if the data
 were from a randomized clinical trial; if these propensities were unknown, they could
 be estimated from the observed data, e.g., using a multinomial logistic regression 
\citep[see also][for related ideas and discussion]{jiang2015doubly, thomas2016data, 
hanna2018importance}. 

An estimating equation for the average utility setting is derived using
a similar strategy to the discounted case.  For each $\pi$ define
the differential value 
\begin{equation*}
\delta({\pi}, \V{s}) \triangleq 
\lim_{N\rightarrow \infty}\mathbb{E}^{\pi}\left[
\sum_{j=1}^{N}\left\lbrace 
U^j - V_{\mathrm{ave}}({\pi})
\right\rbrace
\bigg| \M{S}^1 = \V{s}
\right],
\end{equation*}
which is well-defined under the regularity conditions provided
in Section 5.  Then it can be shown \cite[e.g.,][]{puterman2014markov,murphy2016batch,liao2019off}
that $V_{\mathrm{ave}}({\pi})$ satisfies the recursion 
\begin{equation}\label{PopBellmanAverage}
0 = \mathbb{E}\left[
\frac{\pi_{A^j}(\M{S}^j)
}{
P(A^j|\M{S}^j)  
}\left\lbrace
U^j - V_{\mathrm{ave}}({\pi}) 
+ \delta({\pi}, \M{S}^{j+1}) - \delta({\pi}, \M{S}^j)
\right\rbrace
\psi(\M{S}^j)
\right],
\end{equation}
for all $j$ and any $\psi:\mathcal{S}\rightarrow \mathbb{R}^e$.  
Let $\mathcal{W} = \left\lbrace \delta(\V{S}; \V{\beta}):\,\,\V{\beta} \in
\mathcal{B}\subseteq \mathbb{R}^{e-1}\right\rbrace$ be a 
class of continuously differentiable maps from $\mathcal{S}$ into
$\mathbb{R}^e$.   
An estimator 
$\widehat{V}_{\mathrm{ave},n}({\pi})$ of $V_{\mathrm{ave}}({\pi})$ 
is obtained by jointly solving the
sample analog of (\ref{PopBellmanAverage}) for
$\V{\beta}$ and $V_{\mathrm{ave}}({\pi})$ with
$\psi(\V{s}) = 
\left\lbrace
1, \nabla_{\V{\beta}}\delta (\V{s};\V{\beta})^{\intercal}\right\rbrace^{\intercal}$ 
so that 
$\widehat{V}_{\mathrm{ave},n}({\pi})$, 
$\widehat{\V{\beta}}_{n}({\pi})$ solve
\begin{equation}\label{SampleBellmanAverage}
0 = \mathbb{P}_{n}\sum_{j=1}^{J-1}\left[
\frac{\pi_{A^j}(\M{S}^j)
}{
P(A^j|\M{S}^j)  
}\left\lbrace
U^j - V_{\mathrm{ave}}({\pi}) 
+ \delta({\pi}, \M{S}^{j+1}; \V{\beta}) - 
\delta({\pi}, \M{S}^j;\V{\beta})
\right\rbrace
\left(
{1 \atop \nabla_{\V{\beta}}\delta (\M{S}^j;\V{\beta})}
\right)
\right].
\end{equation}
The estimated optimal regime is thus given by
$\widehat{{\pi}}_{\mathrm{ave}, n} = \arg\max_{{\pi}\in\Pi}
\widehat{V}_{\mathrm{ave},n}({\pi}).$

\begin{remark}
The remainder of this manuscript is focused on constructing the 
transformed process and examining the theoretical and empirical
properties of the foregoing two estimators when applied to
 the transformed data.  However, these are but two
 of many possible methods for estimating an optimal regime with MDPs;
 these were chosen because they have been used previously 
 in clinical applications and, furthermore, are simple, extensible, and amenable to statistical inference
 \cite[for alternative approaches see][and references therein]{szepesvari2010algorithms,powell2007approximate,sutton1998reinforcement}.
\end{remark}

\section{Estimation of sufficient summary functions}\label{hmm}
Recall that the sufficient summary functions are given by
$f^j(\M{H}^j) = \left(\M{B}^j, \M{X}^j\right)$ for $j\ge 1$.  
As $\M{X}^j$ is observed,
constructing an estimator of $f^j$ is tantamount to constructing
an estimator of $\M{B}^j$, the conditional distribution of the latent
state given history $\M{H}^j$.   We develop an estimator of 
$\M{B}^j$ under the assumption that the observables, 
$\M{X}^j$, evolve under a latent-state-dependent autoregressive process.
This choice is motivated by the clinical theory underpinning 
bipolar disorder as well as its robustness and utility in 
modeling chronic illness \citep[for additional discussion on time series and mechanistic models for bipolar disorder, see][and references therein]{daugherty2009mathematical,bonsall2011nonlinear,moore2012forecasting,moore2014mood,bonsall2015bipolar,holmes2016applications}.

We assume that the latent state $M(t)$ follows
a homoegenous Markov process the dynamics of which  are described 
by the transition rate matrix  $Q(a) \triangleq \left\lbrace
q_{k,\ell}(a)
\right\rbrace_{k,\ell=1,\ldots,K}\in\mathbb{R}^{K\times K}$ 
for each $a\in\left\lbrace 1,\ldots, L\right\rbrace$, where 
\begin{eqnarray*}
  q_{k,k}(a)&\triangleq& -\lim_{t\to0^+}
t^{-1}
  P\{ M(T^{j+1})\ne k | T^{j+1}-T^j=t, M(T^j)=k, A^j=a\}, \\
  q_{k,\ell}(a) &\triangleq&\,\,\,\,\,\, \lim_{t\to0^+}
  t^{-1} P\{ M(T^{j+1})=\ell | T^{j+1}-T^j=t, M(T^j)=k, A^j=a\}, \, k\ne \ell, 
\end{eqnarray*}
from which it can be seen that $q_{k,k}(a) = -\sum_{\ell\ne k}
q_{k,\ell}(a)$ for $k=1,\ldots, K$
\citep[see][]{liu2015efficient}.   The transition rate matrix,
also known as the infinitesimal generator \citep[e.g.,][]{pyke1961markova,pyke1961markovb,albert1962estimating}, induces the
following transition probabilities
\begin{equation*}
P\{M(t')=\ell | M(t)=k, A=a\}=[\exp\{(t'-t)\cdot Q(a)\}]_{k,\ell},
\end{equation*}
for $t' > t$ and $k,\ell=1,\ldots, K$. 
We posit parametric models for the dynamics
of the observed data and assume that these models have densities
 of the following form:
the density of $\M{X}^1$ given $M(T^1) = m^1$ is 
$p_{\M{X}^1|M(T^1)}(\V{X}^1|m^1;\M{\theta})$, which is indexed
by $\M{\theta} \in \V{\Theta}$, and the density of
$\M{X}^j$ given 
$M(T^j) = m^j$ 
and  $\M{X}^{j-1}=\V{X}^{j-1}$
is 
$p_{\M{X}^j|M(T^j), \M{X}^{j-1}}(\V{x}^j|m^j, \V{X}^{j-1}; \M{\gamma})$,
which is indexed by $\V{\gamma} \in \M{\Gamma}$. 
For example, a Gaussian autoregressive model with linear mean models takes the form:
\begin{eqnarray*}
p_{\M{X}^1|M(T^1)}(\V{X}^1|m^1;\M{\theta}) &\propto&
|\M{\Sigma}_{m^1}|^{-1/2}
\exp\biggl\{ -\frac{1}{2}(\V{x}^1-\V{\mu}_{m^1})
^\intercal\M{\Sigma}_{m^1}^{-1}(\V{x}^1-\V{\mu}_{m^1}) \biggr\},
\end{eqnarray*}
where $\V{\theta} = \left\lbrace 
(\V{\mu}_{m}, \M{\Sigma}_{m})
\right\rbrace_{m=1,\ldots,K}$
are unknown parameters, and 
\begin{multline*}
p_{\M{X}^j|M(T^j), \M{X}^{j-1}}(\V{x}^j|m^j, \V{X}^{j-1}; \M{\gamma})
\propto\\ 
|\M{\Sigma}_{m^j}|^{-1/2}
\exp\left\lbrace
-\frac{1}{2}(\V{x}^j-
\V{\Psi}_{m^{j}}\V{x}^{j-1})^{\intercal}
\M{\Sigma}_{m^{j}}^{-1}
(\V{x}^j- \V{\Psi}_{m^{j}}\V{x}^{j-1})
\right\rbrace,
\end{multline*}
where $\M{\gamma} = \left\lbrace
(\M{\Psi}_{m}, \M{\Sigma}_{m})
\right\rbrace_{m=1,\ldots,K}.$  We use this model in our 
simulation experiments and application to the data from the STEP-BD
trial.  

Let $\V{\varrho} = (\V{\theta}, \V{\gamma}) \in \V{\Theta}\times \V{\Gamma}$ denote the unknown parameters
indexing the latent Markov process.  It can be seen that
$\M{B}^j$ is determined by $\M{H}^j$ and $\V{\varrho}$, i.e.,
$\M{B}^j = \V{b}^j(\M{H}^j, \V{\varrho})$ where $\V{b}^j$ is a deterministic
map from $\mathrm{dom}\,\M{H}^j \times (\V{\Theta}\times \V{\Gamma})$  into $\mathcal{D}$, the 
$L$-dimensional probability simplex.  
We construct an estimator
$\widehat{\V{\varrho}}_n$ via maximum likelihood implemented using
the forward-backward algorithm \citep[for a review see][]{rabiner1989tutorial}
and subsequently compute the plug-in estimator 
$\widehat{\M{B}}_n^j = \V{b}^j(\M{H}^j, \widehat{\V{\varrho}}_n)$ so that   
$\widehat{\M{S}}_n^j  = (\widehat{\M{B}}_n^j, \M{X}^j)$. The preceding estimator
is used to convert i.i.d. trajectories of the form
$\left\lbrace \left(T_i^j, A_i^j, \M{X}_i^j\right)\right\rbrace_{j=1}^{J_i}$ for 
$i=1,\ldots, n$ into trajectories drawn from an (approximate) homogeneous
MDP $\left\lbrace (\widehat{\M{S}}_{n,i}^j, A_{i}^j, \widehat{U}_{n,i}^j) \right\rbrace_{j=1}^{J_i}$ for
$i=1,\ldots, n$, which can then be used with the estimators of an optimal 
regime described in the previous section.


\section{Theoretical properties}
We establish consistency and asymptotic normality for the
estimated optimal regime constructed by solving estimating
equations as described in Section 3 applied to the transformed data.
For simplicity, in Sections 5.2 and 5.3 we assume that the
class of regimes ${\Pi}$ is finite. 
However, this assumption is not limiting as
given an arbitrary $\eta >0$
one can approximate any separable collection of regimes, $\widetilde{{\Pi}}$, by a
finite mesh, ${\Pi}$, so that $\sup_{{\pi}\in{\Pi}}V({\pi})$ 
is within $\eta$ of $\sup_{\widetilde{{\pi}}\in\widetilde{{\Pi}}}V(\widetilde{{\pi}})$  
\citep[see][for additional discussion]{zhang2018interpretable}. 
We illustrate this approach in Section 5.4 when we derive confidence sets
for the value of the optimal regime within a parametric 
class of regimes.

\subsection{Consistency of the  estimated state probabilities}
Consistency of the estimated latent state distribution is central to
characterizing the large sample behavior of estimators of the
optimal treatment regime constructed by solving the MDP estimating equations
of Section 3.  Consistency follows from existing results on 
maximum likelihood for latent Markov models and the continuous
mapping theorem.  
We make the following assumptions.  
\bigskip
\noindent
\begin{enumerate}
\item [(B1)] Both the time process $(T^j: j\in\mathbb{N})$ and 
the number of time points $J$ are
  independent of the latent process $\left\lbrace M(t)\, :\, t \ge 0\right\rbrace$.
\item [(B2)] The true parameter vector $\V{\varrho}^*$ is an interior point of
  $\M{\Theta}\times \M{\Gamma}$, where $\M{\Theta}\times\M{\Gamma}$ is a compact subset of
  $\mathbb{R}^{\dim\,\V{\varrho}}$. 
\item [(B3)] For all $m^1\in\{1,\ldots,K\}$, $P\{M(T^1)=m^1\}>0$.
\item [(B4)] There exist measures $\upsilon_1, \upsilon_2$ on $\mathcal{X}$ which are bounded
away from zero with
\begin{align*}
& p_{\M{X}^1|M(T^1)}(\cdot|m^1)\geq\upsilon_1(\cdot)
\textrm{ for all }m^1\in\{1,\ldots,K\}, \\
& p_{\M{X}^j|M(T^j),\M{X}^{j-1}}(\cdot|m^j, \V{x}^{j-1})
\geq\upsilon_2(\cdot)
\textrm{ for all }m^j\in\{1,\ldots,K\}, \V{x}^j \in \mathcal{X}, j > 1,
\end{align*}
where $\mathcal{X} = \mathrm{dom}\,\M{X}^j\,, j \ge 1$. 
\item [(B5)] For each $\V{\varrho}\in \M{\Theta}\times\M{\Gamma}$, the transition
  kernel indexed by $\V{\varrho}$ is stationary, Harris recurrent, and
  aperiodic \citep[see][for additional discussion of this assumption and its implications]{athreya2006measure,meyn2012markov}. 
\item [(B6)] The transition kernel is continuous in $\pmb{\varrho}$ in
  an open neighborhood of $\pmb{\varrho}^*$.
\item [(B7)] The latent Markov process is identifiable up to label switching of the latent states
\citep[see][for discussions of label-switching]{allman2009identifiability,gassiat2013finite}.
\item [(B8)] The log likelihood is twice continuously differentiable in $\V{\varrho}$ and 
the Fisher information  $\M{I}(\V{\varrho})$ is positive definite
in an open neighborhood of $\pmb{\varrho}^*$ 
\citep[see][for equivalent assumptions]{bickel1998asymptotic,jensen1999asymptotic,douc2004asymptotic}.
\item [(B9)] For any $\V{\varrho}_1,\V{\varrho}_2\in\M{\Theta}\times\M{\Gamma}$, 
$\|\V{b}^j(\V{h}^j,\V{\varrho}_1)-\V{b}^j(\V{h}^j,\V{\varrho}_2)\|\leq g^j(\V{h}^j)\|\V{\varrho}_1-\V{\varrho}_2\|,$
for some integrable function $g^j: \mathrm{dom}\,\M{H}^j\to\mathbb{R}$, $j\geq1$.
\end{enumerate}
\noindent 
The preceding assumptions are relatively mild and standard in hidden Markov models. 
Assumption (B1) ensures that the distribution of the visit times factors out of the likelihood
for $\V{\varrho}$, i.e., the time process and latent process do not share parameters.  
Assumptions (B2)-(B8) ensure that the model is well-defined and that the maximum likelihood 
estimators are regular 
 \citep[][]{leroux1992maximum,bickel1998asymptotic,jensen1999asymptotic,le2000exponential,douc2001asymptotics,douc2004asymptotic}. 
Consistency and asymptotic normality of the maximum likelihood estimators in
general autoregressive hidden Markov models have been 
established under the preceding conditions
\citep[][]{douc2004asymptotic}.
Moreover, we show that Assumption (B9) holds for the Gaussian autoregressive
hidden Markov model in the Supplementary Material, 
which ensures the class is Donsker and thus 
consistency 
of the estimated state probabilities follows immediately. 
\begin{lemma}\label{lemmahmm}
Assume (A1), (B1) - (B8), as $n\to\infty$:
\begin{align*}
&\sqrt{n}(\widehat{\V{\varrho}}_n-\V{\varrho}^*) \rightsquigarrow\mathcal{N}\{\V{0}, \M{I}(\V{\varrho}^*)^{-1}\},\\
&\sqrt{n}\{\V{b}^j(\V{h}^j; \widehat{\V{\varrho}}_n) - 
\V{b}^j(\V{h}^j; \V{\varrho}^*)\} \rightsquigarrow \mathcal{N}\{\V{0}, 
\nabla_{\V{\varrho}}\V{b}^j(\V{h}^j; \V{\varrho}^*) \M{I}(\V{\varrho}^*)^{-1}\nabla_{\V{\varrho}}\V{b}^j(\V{h}^j; \V{\varrho}^*)^\intercal\},
\end{align*}
for each $\V{h}^j\in \mathrm{dom}\,\M{H}^j$.
Furthermore, if (B9) holds, then for each fixed $j=1,\dots,J$, as $n\to\infty$,
\begin{align*}
&\sup_{\V{h}^j\in\mathrm{dom}\,\M{H}^j}\big|
\V{b}^j(\V{h}^j, \widehat{\V{\varrho}}_n)-
\V{b}^j(\V{h}^j, \pmb{\varrho}^*)\big|\overset{p}{\to}0.
\end{align*}
\end{lemma}

\subsection{Asymptotic properties in the discounted utility setting}

We consider linear working models for the state-value function 
$\nu(\V{s};\V{\alpha})=\phi(\V{s})^\intercal\V{\alpha}$,
where $\phi$ is a finite-dimensional set of basis functions; these 
basis functions might comprise custom features informed by domain
expertise as well as nonlinear expansions such as 
b-splines or radial basis functions.
Using this functional form, the population-level estimating equation 
for the state value-function, i.e., (\ref{popnEstEqVLearning})
from Section \ref{dtr}, is given by 
\begin{equation*}
\Lambda_{\mathrm{dis}}({\pi}, \V{\alpha})\triangleq
\mathbb{E} \sum_{j=1}^{J-1} 
\left[
\frac{
  \pi_{A^j}(\M{S}^j)
}{
  P(A^j\big|\M{S}^j)
}
\left\lbrace
U^j + \gamma \phi(\M{S}^{j+1})^{\intercal}\V{\alpha}
- \phi(\M{S}^j)^{\intercal}\V{\alpha}
\right\rbrace  \phi(\M{S}^j)
\right];
\end{equation*}
let $\V{\alpha}^*({\pi})$ denote the solution
to $\Lambda_{\mathrm{dis}}({\pi}, \V{\alpha}) = 0$.  
The sample analog using
the estimated states  
is thus 
\begin{equation*}
\widehat{\Lambda}_{\mathrm{dis},n}({\pi}, \V{\alpha})\triangleq
\mathbb{P}_n\sum_{j=1}^{J-1} 
\left[
\frac{
  \pi_{A^j}\left(\widehat{\M{S}}_{n}^j\right)
}{
  P\left(A^j\big|\widehat{\M{S}}_{n}^j\right)
}
\left\lbrace
\widehat{U}_n^j + \gamma 
\phi\left(\widehat{\M{S}}_{n}^{j+1}\right)^{\intercal}\V{\alpha}
-\phi\left(\widehat{\M{S}}_{n}^j\right)^{\intercal}
\V{\alpha}
\right\rbrace \phi\left(\widehat{\M{S}}_{n}^j\right)
\right],
\end{equation*}
where $\widehat{U}_n^j = \mathcal{U}(\widehat{\M{S}}_n^j, A^j)$;
let $\widehat{\V{\alpha}}_n({\pi})$  denote a solution
to $\widehat{\Lambda}_{\mathrm{dis},n}({\pi}, \V{\alpha}) = 0$. 
For 
linear estimators, such a root always exists, however, below we 
require the weaker condition that $\widehat{\V{\alpha}}_{n}({\pi})$
is an approximate root.  
Let $||\cdot||_F$ denote the Frobenius norm. We make the following assumptions.  
\noindent
\begin{enumerate}
\item [(C1)] For each ${\pi}\in {\Pi}$, 
  $\V{\alpha}^*({\pi})$ solves $\Lambda_{\mathrm{dis}}({\pi},\V{\alpha})=0$,
  where $\V{\alpha}^*({\pi})$ is an interior point of
  $\M{\Omega}$ and $\M{\Omega}$ is a compact subset of
  $\mathbb{R}^{\dim\,\V{\alpha}}$. 
  \item[(C2)] For each ${\pi}\in {\Pi}$, there exists a sequence of $\widehat{\V{\alpha}}_n({\pi})\in\M{\Omega}$
  such that $\widehat{\Lambda}_{\mathrm{dis},n}\{{\pi},\widehat{\V{\alpha}}_n({\pi})\}=o_p(n^{-1/2})$.
  \item[(C3)] Define $V_{\mathrm{dis}}^R({\pi}) \triangleq \int \phi(\V{s})^{\intercal}\V{\alpha}^*({\pi}) dR(\V{s})$,
  which attains its supremum at ${\pi}^*_{\mathrm{dis}} \in {\Pi}$.
\item [(C4)] There exists a sequence of $\widehat{{\pi}}_{\mathrm{dis},n}^R\in {\Pi}$
such that
$\widehat{V}_{\mathrm{dis},n}^R(\widehat{{\pi}}_{\mathrm{dis},n})\geq 
\sup_{\pi\in\Pi}\widehat{V}_{\mathrm{dis},n}^R({\pi}) - o_p(1).$
\item [(C5)] There exists a constant $c>0$, such that 
$$\V{\omega}^{\intercal}  \mathbb{E}\biggl[\frac{\pi_{A^j}(\M{s}^j)}{P(A^j|\M{S}^j)}
\phi(\M{S}^j)\biggl\{ \phi(\M{s}^j - \gamma\phi(\M{s}^{j+1}) \biggr\}^{\intercal} 
\biggr] \V{\omega}\geq c\|\V{\omega}\|_2^2,$$ 
for all $j\geq1$ and $\V{\omega}\ne\V{0}$.
\item [(C6)]  
$\phi: \mathcal{S}\to\mathbb{R}^d$ is uniformly continuous, where $\mathcal{S} = \mathrm{dom}\,\M{S}^j\,, j = 1,\ldots, J$, is compact, and 
$J$ is finite almost surely.
Furthermore, $\mathbb{E}||\M{S}^j||^2 < \kappa$ for some $\kappa > 0$ and all $j\geq 1$.  
\item [(C7)] 
For each $a\in\mathcal{A}$, $\V{s} \in \mathcal{S}$, $j\geq1$, $P\left(A^j=a|\M{S}^j=\V{s}\right)\ge \epsilon$,
for some $\epsilon>0$. 
\item [(C8)] For each $j=1,\ldots,J$, ${\pi}\in{\Pi}$, define
\begin{equation*}
\M{G}^j_{\mathrm{dis}}\left(
{\pi}, \V{h}^j, a^j; \V{\varrho}
\right) \triangleq \frac{
  \pi_{a^j}\left\lbrace 
  b^j(\V{h}^j; \V{\varrho}) 
  \right\rbrace
  }{
  P\left\lbrace a^j|b^j(\V{h}^j;\V{\varrho})
  \right\rbrace
  }
  \mathcal{U}\left\lbrace
    b^j\left(\V{h}^j; \V{\varrho}\right), a^j
  \right\rbrace
  \phi\left\lbrace 
    b^j\left(
        \V{h}^j; \V{\varrho}
       \right) 
  \right\rbrace.
\end{equation*}
There exists a linear operator $\M{W}^j_{\mathrm{dis}}({\pi}, \V{H}^j, a^j; \V{\varrho})$
such that $\mathbb{E}||\M{W}^j_{\mathrm{dis}}({\pi}, \M{H}^j, A^j; \V{\varrho})||_F < \infty$ 
and, for all $\V{h}^j\in\mathrm{dom}\,\M{H}^j$ and $a^j\in\mathcal{A}$, the following expansion holds 
\begin{equation*}
\M{G}^j_{\mathrm{dis}}\left({\pi}, \V{H}^j, a^j; \widehat{\V{\varrho}}_n\right) - 
\M{G}^j_{\mathrm{dis}}\left({\pi}, \V{H}^j, a^j; \V{\varrho}^*\right) = 
\M{W}^j_{\mathrm{dis}}({\pi}, \V{H}^j, a^j; \V{\varrho}^*)
(\widehat{\V{\varrho}}_n-\V{\varrho}^*)+ o_p(n^{-1/2}).
\end{equation*}
\end{enumerate}
These conditions are standard for Z-estimators \citep[][]{van1996weak,kosorok2008introduction}.
 Conditions (C1), (C2), (C6), and (C7) are used to establish the consistency of $\widehat{\V{\alpha}}_n({\pi})$,
while the addition of (C5) and (C8) are used to establish asymptotic normality.
A sufficient condition for (C8) is that
$\M{G}^j_{\mathrm{dis}}({\pi}, \V{H}^j, a^j; \V{\varrho})$ 
is almost everywhere differentiable in $\V{\varrho}$
in which case $\M{W}^j_{\mathrm{dis}}({\pi}, \V{H}^j, a^j; \V{\varrho})$ can be chosen
 to be the gradient operator.
We use (C3) and (C4) to show  
$\widehat{V}_{\mathrm{dis},n}(\widehat{{\pi}}_{\mathrm{dis},n})\overset{p}{\to} V_{\mathrm{dis}}({\pi}^*_{\mathrm{dis}})$,
which is a weaker but more general result than 
$\widehat{{\pi}}_{\mathrm{dis},n}\overset{p}{\to}{\pi}^*_{\mathrm{dis}}$.
Convergence of $\widehat{{\pi}}_{\mathrm{dis},n}$ generally requires 
that ${\pi}^*_{\mathrm{dis}}$ is a unique and well-separated maximizer
of $V_{\mathrm{dis}}({\pi})$, which need not hold for some commonly
used classes of regimes \citep[see][]{zhang2018interpretable}.  
 

\begin{theorem}\label{theoremdis1}
Assume (A1), (C1) - (C7), and that ${\Pi}$ is finite. Then as $n\to\infty$:
\begin{enumerate}
\item for any fixed regime ${\pi}$, $\widehat{\V{{\alpha}}}_{n}({\pi})\overset{p}{\to}\V{\alpha}^*({\pi})$;  
\item $\widehat{V}_{\mathrm{dis},n}(\widehat{{\pi}}_{\mathrm{dis},n})\overset{p}{\to} 
V_{\mathrm{dis}}({\pi}^*_\mathrm{{dis}})$. 
\end{enumerate}
\end{theorem}

To define the limiting distribution of the estimated optimal value  we make use of the following quantities: 
\begin{align*}
C_{1}({\pi})\triangleq &\mathbb{E}\biggl[  \sum_{j=1}^{J-1} \frac{ \pi_{A^j}(\M{S}^j)}{P(A^j|\M{S}^j)}
\phi(\M{s}^j)\biggl\{ \phi(\M{s}^j) -  \gamma\phi(\M{s}^{j+1})\biggr\}^{\intercal} \biggr],\\[3pt]
\widehat{C}_{1,n}({\pi})\triangleq &
\mathbb{P}_n\biggl[  
\sum_{j=1}^{J-1} 
\frac{ 
  \pi_{A^j}(\widehat{\M{s}}_{n}^j)
  }{
  P(A^j|\widehat{\M{s}}_{n}^j)}
\phi(\widehat{\M{s}}_{n}^j)
\biggl\{ \phi(\widehat{\M{s}}_{n}^j) -  
\gamma\phi(\widehat{\M{s}}_{n}^{j+1})
\biggr\}^{\intercal} 
\biggr],\\[3pt]
C_{2}({\pi},\widetilde{{\pi}})\triangleq&\mathbb{E}\biggl[ \sum_{j=1}^{J-1} 
\frac{\pi_{A^j}(\M{s}^j)\widetilde{\pi}_{A^j}(\M{s}^j)}{P^2(A^j|\M{s}^j)}  
\biggl\{ U^j + \gamma \phi(\M{s}^{j+1})^{\intercal} \V{\alpha}^*({\pi})
- \phi\{\M{s}^j\}^{\intercal}\V{\alpha}^*({\pi})  \biggr\} \\
& \quad\quad
\biggl\{ U^j + \gamma \phi(\M{s}^{j+1})^{\intercal}\V{\alpha}^*(\widetilde{{\pi}})
- \phi(\M{s}^j)^{\intercal}\V{\alpha}^*(\widetilde{{\pi}})  \biggr\}\phi(\M{s}^j)\phi(\M{s}^j)^{\intercal} \biggr],\\[3pt]
\widehat{C}_{2,n}({\pi}, \widetilde{{\pi}}) \triangleq & 
\mathbb{P}_n\biggl[
 \sum_{j=1}^{J-1}  
  \frac{\pi_{A^j}(\widehat{\M{s}}_{n}^j)\widetilde{\pi}_{A^j}(\widehat{\M{s}}_{n}^j)
  }{
  P^2(A^j|\widehat{\M{s}}_{n}^j)
  }  
\biggl\{ 
\widehat{U}_{n}^j + \gamma \phi(\widehat{\M{s}}_{n}^{j+1})^{\intercal}\widehat{\V{\alpha}}_n({\pi})
- \phi(\widehat{\M{s}}_{n}^j)^{\intercal}\widehat{\V{\alpha}}_n({\pi})  
\biggr\} \\
&\quad\quad
\biggl\{ 
\widehat{U}_{n}^j + \gamma \phi\left(\widehat{\M{s}}_{n}^{j+1}\right)^{\intercal}\widehat{\V{\alpha}}_n(\widetilde{{\pi}})
- \phi(\widehat{\M{s}}_{n}^j)^{\intercal}\widehat{\V{\alpha}}_n(\widetilde{{\pi}})  
\biggr\}
\phi(\widehat{\M{s}}_{n}^j)
\phi(\widehat{\M{s}}_{n}^j)^\intercal \biggr],\\
C_3({\pi})\triangleq & \mathbb{E}\biggl[ 
\sum_{j=1}^{J-1}\M{W}^j_{\mathrm{dis}}({\pi}, \M{H}^j, A^j; \V{\varrho}^*)
\biggr],\\
\widehat{C}_{3,n}({\pi})\triangleq & \mathbb{P}_n\biggl[ 
\sum_{j=1}^{J-1} \M{W}^j_{\mathrm{dis}}({\pi}, \M{H}^j, A^j; \widehat{\V{\varrho}}_n)
\biggr].
\end{align*}

\begin{theorem}\label{theoremdis2}
Assume (A1), (C1) - (C8), and that ${\Pi}$ is finite.  
The following results hold as $n\rightarrow \infty:$ 
\begin{enumerate}
\item  
$\sqrt{n}\{\widehat{V}_{\mathrm{dis},n}^R({\pi}) - V_{\mathrm{dis}}^R({\pi})\}\rightsquigarrow \mathbb{B}^R({\pi}),$
where $\mathbb{B}^R({\pi})$ is a mean zero Gaussian process indexed by ${\pi}\in\Pi$ with covariance 
$$\mathbb{E}\{\mathbb{B}^R({\pi})\mathbb{B}^R(\widetilde{{\pi}})\}=
\biggl[\int\phi(\V{s}) dR(\V{s})\biggr]^\intercal C_{1}^{-1}({\pi})
[C_{2}({\pi},\widetilde{{\pi}})+C_3({\pi}) \M{I}^{-1}(\V{\varrho}^*) C_3^\intercal(\widetilde{{\pi}})] 
C_{1}^{-\intercal}(\widetilde{{\pi}})\biggl[\int\phi(\V{s}) dR(\V{s})\biggr];$$  
\item  
$\sqrt{n}\widehat{\sigma}^{-1/2}_{\mathrm{dis},n}({\pi})
\left\lbrace
\widehat{V}_{\mathrm{dis},n}^R({\pi})-V_{\mathrm{dis}}^R({\pi}) 
\right\rbrace 
\rightsquigarrow\mathcal{N}(0,1),$
where $$\widehat{\sigma}^{-1/2}_{\mathrm{dis},n}({\pi})=\biggl[\mathbb{P}_n\{\phi(\V{s})\}^{\intercal}\widehat{C}_{1,n}^{-1}({\pi})
[\widehat{C}_{2,n}({\pi},{\pi}) +
\widehat{C}_{3,n}({\pi})\widehat{\M{I}}_n^{-1}(\widehat{\V{\varrho}}_n)\widehat{C}_{3,n}^\intercal({\pi})]
\widehat{C}_{1,n}^{-\intercal}({\pi})\mathbb{P}_n\{\phi(\V{s})\}\biggr]^{1/2}.$$
\end{enumerate}
\end{theorem}



\subsection{Asymptotic properties in the average utility setting}
We derive the limiting distribution of the value function under 
a linear working model for the differential value  
$\delta(\V{s}, \V{\beta}) = \phi(\V{s})^{\intercal}\V{\beta},$ where
$\phi$ is a vector of features constructed from $\V{s}$ as in the
preceding section.  Further, let $\V{\zeta} = (v,\V{\beta}^{\T})^{\T}$ denote a generic vector
in $\mathbb{R}^{e}$, and define 
$\psi_{1}(\V{s}) \triangleq \left\lbrace 1, \phi^{\T}(\V{s})\right\rbrace^{\T}$
and $\psi_2(\V{s}, \widetilde{\V{s}}) \triangleq \left\lbrace 
1, \phi^{\T}(\V{s}) - \phi^{\T}(\widetilde{\V{s}})
\right\rbrace^{\T}$.  The population estimating equation for the average
utility, i.e., equation (\ref{PopBellmanAverage}) in Section \ref{dtr},  under the posited model   
is
\begin{equation*}
\Lambda_{\mathrm{ave}}\left( 
{\pi},\V{\zeta}
\right)
  \triangleq \mathbb{E}\sum_{j=1}^{J-1} \biggl[
\frac{\pi_{A^j}(\M{s}^j)}{P(A^j|\M{s}^j)}
\left\lbrace
U^j 
-\psi_2(\M{s}^j,\M{s}^{j+1})^{\intercal}\V{\zeta}
\right\rbrace 
\psi_1(\M{s}^j)\biggr];
\end{equation*}
define $\V{\zeta}^*({\pi}) = \left\lbrace V_{\mathrm{ave}}^*(\pi), \V{\beta}^{*\T}\right\rbrace^{\T}$ as the solution to 
$\Lambda_{\mathrm{ave}}\left( 
{\pi},\V{\zeta}
\right) = 0.$ The sample analog is 
\begin{equation*}
\widehat{\Lambda}_{\mathrm{ave},n}
\left(
{\pi},\V{\zeta}
\right)  
 \triangleq \mathbb{P}_n \sum_{j=1}^{J-1} \biggl[
\frac{\pi_{A^j}(\widehat{\M{s}}_{n}^j)}{P(A^j|\widehat{\M{s}}_{n}^j)}\{\widehat{U}_{n}^j 
-\psi_2(\widehat{\M{s}}_{n}^j,\widehat{\M{s}}_{n}^{j+1})^{\intercal}\V{\zeta}
\}\psi_1(\widehat{\M{s}}_{n}^j)\biggr],
\end{equation*}
where $\widehat{U}_n^j = \mathcal{U}\left(\widehat{\M{S}}_n^j, A^j\right)$;
define $\widehat{\V{\zeta}}_{n}({\pi}) = 
\left\lbrace \widehat{V}_{\mathrm{ave},n}(\pi), \widehat{\V{\beta}}_{n}^{\T}\right\rbrace^{\T}$ as the solution to
$\widehat{\Lambda}_{\mathrm{ave},n}
\left( {\pi},\V{\zeta}
\right) = 0$. As in the discounted setting,  one can always 
find an exact root to the sample estimating equation 
under a linear model; however, the theory permits approximate
roots as well.

To study the large sample properties of $\widehat{\V{\zeta}}_{n}({\pi})$
we make use of the following regularity conditions. 
\noindent
\begin{enumerate}
\item [(D1)] There exists a measure $\upsilon$ on $\mathcal{S}$ which is bounded
away from zero with
$$p_{\M{S}^j|\M{S}^{j-1}, A^{j-1}}(\cdot | \V{s}, a) \geq 
\upsilon(\cdot)\textrm{ for all }\V{s}\in\mathcal{S}, a\in\mathcal{A}, $$
where $p_{\M{S}^j|\M{S}^{j-1}, A^{j-1}}$ denotes the density of $\M{S}^j$ given
$\M{S}^{j-1}$ and $A^{j-1}$.
\item [(D2)] For all ${\pi}\in {\Pi}$ and $\V{s}\in\mathcal{S}$,
$$\lim\sup_{N\to\infty} \mathbb{E}^{\pi} 
\left(
 \frac{1}{N}\sum_{j=1}^N U^{j} | \M{s}^1=\V{s}
 \right) 
 = \lim\inf_{N\to\infty} \mathbb{E}^{\pi} 
 \left( 
 \frac{1}{N}\sum_{j=1}^N U^{j} | \M{s}^1=\V{s}
 \right).$$
 \item[(D3)] For all $\V{s}\in\mathcal{S}$, 
$ \lim_{\gamma\uparrow1}\biggl[\nu_{\mathrm{dis}}({\pi},\V{s}) - 
V_{\mathrm{ave}}({\pi})/(1-\gamma) \biggr]= O(1).$
\item [(D4)] For each ${\pi}\in {\Pi}$, 
  $\V{\zeta}^*({\pi})$ solves $\Lambda_{\mathrm{ave}}({\pi},\V{\zeta})=0$,
  where $\V{\zeta}^*({\pi})$ is an interior point of
  $\mathcal{Z}$, and $\mathcal{Z}$ is a compact subset of
  $\mathbb{R}^{\dim\,\V{\zeta}}$. 
  \item[(D5)] For each ${\pi}\in {\Pi}$, there exists a sequence of $\widehat{\V{\zeta}}_n({\pi})\in\mathcal{Z}$
  such that $\widehat{\Lambda}_{\mathrm{ave},n}\{{\pi},\widehat{\V{\zeta}}_n({\pi})\}=o_p(n^{-1/2})$.
  \item [(D6)] $V_{\mathrm{ave}}({\pi})$ attains its supremum over ${\Pi}$ at ${\pi}^*_{\mathrm{ave}}$.
  \item [(D7)] There exists a sequence $\widehat{{\pi}}_{\mathrm{ave},n}\in {\Pi}$
 such that
$\widehat{V}_{\mathrm{ave},n}(\widehat{{\pi}}_{\mathrm{ave},n})\geq 
\sup_{\pi\in\Pi}\widehat{V}_{\mathrm{ave},n}({\pi}) - o_p(1).$
\item [(D8)] There exists a constant $c>0$, such that 
$$\V{w}^{\T}  \mathbb{E}
\left\lbrace
\frac{\pi_{A^j}(\M{s}^j)}{P(A^j|\M{S}^j)}
\psi_1(\M{S}^j)\psi_2(\M{s}^j,\M{s}^{j+1})^{\T}
\right\rbrace 
\V{w}\geq c\|\V{w}\|_2^2,$$ 
for all $j>0$ and $\V{w}\ne\V{0}$.
\item [(D9)] For each $j=1,\ldots,J$, ${\pi}\in{\Pi}$, define
\begin{equation*}
\M{G}_{\mathrm{ave}}^j\left(
{\pi}, \V{h}^j, a^j; \V{\varrho}
\right) \triangleq 
\frac{
  \pi_{a^j}\left\lbrace 
  b^j\left(\V{h}^j;\V{\varrho}\right)
  \right\rbrace 
}{
  P\left\lbrace 
  a^j| b^j\left(\V{h}^j; \V{\varrho}\right)
  \right\rbrace 
}\mathcal{U}\left\lbrace
  b^j\left(\V{h}^j; \V{\varrho}\right), a^j
  \right\rbrace \psi_1\left\lbrace
    b^j\left(\V{h}^j; \V{\varrho}\right)
  \right\rbrace.
\end{equation*}
There exists a linear operator $\M{W}^j_{\mathrm{ave}}({\pi}, \V{H}^j, a^j; \V{\varrho})$
such that,  $\mathbb{E}||\M{W}_{\mathrm{ave}}^j({\pi}, \M{H}^j, A^j; \V{\varrho}^*)||_{F} < \infty$, 
and for all $\V{h}^j\in\mathrm{dom}\,\M{H}^j$ and $a^j\in\mathcal{A}$,
$$\M{G}^j_{\mathrm{ave}}\left(
{\pi}, \V{H}^j, a^j;\widehat{\V{\varrho}}_n
\right) - 
\M{G}^j_{\mathrm{ave}}\left(
{\pi}, \V{H}^j, a^j;\V{\varrho}^*
\right) = 
\M{W}^j_{\mathrm{ave}}({\pi}, \V{H}^j, a^j; \V{\varrho}^*)
(\widehat{\V{\varrho}}_n-\V{\varrho}^*)+ o_p(n^{-1/2}).$$
\end{enumerate}
Assumption (D1) is a common  condition 
in the average utility MDP setting 
 \citep[see][for variants of this assumption]{yamada1975duality,kurano1986markov,cavazos1988necessary,hernandez1991recurrence}.
 This assumption ensures that there is a nonzero transition density
 from any starting state  to any other state  under all feasible regimes.
A consequence is that $V_{\mathrm{ave}}({\pi})$ does not depend on the starting state. 
Assumption (D2) guarantees the existence of $V_{\mathrm{ave}}({\pi})$ as the limit of the expected average potential utility
for all ${\pi}\in\Pi$. 
Assumption (D3) requires the system dynamics be such that as 
$\gamma\uparrow 1$, $\nu_{\mathrm{dis}}({\pi}, \V{s})$ behaves like
the expected total utility starting from $\V{s}$ while 
$V_{\mathrm{ave}}({\pi})/(1-\gamma)$ behaves like the 
discounted total utility averaging across initial states.
The remainder of the assumptions are standard regularity 
assumptions for Z-estimators \citep[][]{kosorok2008introduction}.

\begin{theorem}\label{theoremavg1}
Assume (A1), (C6) - (C7), (D1) - (D8), and that ${\Pi}$ is finite. Then as $n\rightarrow \infty:$
\begin{enumerate}
\item For any fixed regime ${\pi}\in\Pi$, 
$\widehat{\V{\zeta}}_{n}({\pi})\overset{p}{\to}\V{\zeta}^*({\pi})$, 
as $n\to\infty$. 
\item $\widehat{V}_{\mathrm{ave},n}(\widehat{{\pi}}_{\mathrm{ave},n})\overset{p}{\to}
V_{\mathrm{ave}}({\pi}^*_{\mathrm{ave}})$ as $n\to\infty$. 
\end{enumerate}
\end{theorem}

\noindent 
The following quantities will be used defining limiting distribution of 
$\widehat{\V{\zeta}}_{n}({\pi})$:
\begin{align*}
D_{1}({\pi})\triangleq &
\mathbb{E}
\left\lbrace  
\sum_{j=1}^{J-1} \frac{ \pi_{A^j}(\M{s}^j)}{P(A^j|\M{s}^j)}
\psi_1(\M{s}^j)\psi_2(\M{s}^j,\M{s}^{j+1})^{\T} 
\right\rbrace,
\\[3pt]
\widehat{D}_{1,n}({\pi})\triangleq 
&\mathbb{P}_n
\left\lbrace
\sum_{j=1}^{J-1}  
\frac{ 
  \pi_{A^j}\left(\widehat{\M{s}}_{n}^j\right)
  }{
  P\left(A^j|\widehat{\M{s}}_{n}^j\right)
  }
\psi_1\left(\widehat{\M{s}}_{n}^j\right)
\psi_2\left(
\widehat{\M{s}}_{n}^j,\widehat{\M{s}}_{n}^{j+1}
\right)^{\T} 
\right\rbrace,\\[3pt]
D_{2}({\pi},\widetilde{\V{\pi}})\triangleq
&\mathbb{E}
\biggl[ 
\sum_{j=1}^{J-1} 
\frac{
  \pi_{A^j}\left(\M{s}^j\right)\widetilde{\pi}_{A^j}\left(\M{s}^j\right)
  }{
  P^2\left(A^j|\M{s}^j\right)
  }  
\biggl\{ 
U^j - \psi_2(\M{s}^j,\M{s}^{j+1})^{\T}\V{\zeta}^*({\pi})  
\biggr\} \\
&\quad\quad
\biggl\{ 
U^j - \psi_2
\left(
\M{s}^j,\M{s}^{j+1}
\right)^{\T}
\V{\zeta}(\widetilde{{\pi}})   
\biggr\}
\psi_1(\M{s}^j)\psi_1(\M{s}^j)^{\T} 
\biggr],\\[3pt]
\widehat{D}_{2,n}({\pi},\widetilde{{\pi}})\triangleq
&\mathbb{P}_n
\biggl[ 
\sum_{j=1}^{J-1} 
\frac{
  \pi_{A^j}
  \left(
  \widehat{\M{s}}_{n}^j
  \right)
  \widetilde{\pi}_{A^j}
  \left(\widehat{\M{s}}_{n}^j\right)
}{
  P^2\left(A^j|\widehat{\M{s}}_{n}^j\right)
}  
\biggl\{ 
\widehat{U}_{n}^j -\psi_2
\left(
\widehat{\M{s}}_{n}^j,\widehat{\M{s}}_{n}^{j+1}
\right)^{\T}
\widehat{\V{\zeta}}_n\left({\pi}\right) 
 \biggr\} \\
& \quad\quad
\biggl\{ \widehat{U}_{n}^j - \psi_2\left(\widehat{\M{s}}_{n}^j,\widehat{\M{s}}_{n}^{j+1}\right)^{\T}\widehat{\V{\zeta}}_n(\widetilde{{\pi}})   \biggr\}
\psi_1\left(\widehat{\M{s}}_{n}^j\right)\psi_1\left(\widehat{\M{s}}_{n}^j\right)^{\T} \biggr],\\
D_3({\pi})\triangleq & \mathbb{E}\biggl[ 
\sum_{j=1}^{J-1}\M{W}^j_{\mathrm{ave}}({\pi}, \M{H}^j, A^j; \V{\varrho}^*)
\biggr],\\
\widehat{D}_{3,n}({\pi})\triangleq & \mathbb{P}_n\biggl[ 
\sum_{j=1}^{J-1} \M{W}^j_{\mathrm{ave}}({\pi}, \M{H}^j, A^j; \widehat{\V{\varrho}}_n)
\biggr].
\end{align*}



\begin{corollary}\label{corollaryavg3}
Assume (A1), (C6) - (C7), (D1) - (D9), and that ${\Pi}$ is finite.
Then for each ${\pi}\in {\Pi}$, as $n\rightarrow \infty$:
\begin{equation*}
\sqrt{n}\{\omega({\pi})\}^{-1/2}\{\widehat{V}_{\mathrm{ave},n}({\pi}) - V_{\mathrm{ave}}({\pi}) \} \rightsquigarrow \mathcal{N}(0,1),
\end{equation*}
where $\omega({\pi})$ is the element at entry $(1,1)$ of 
$$\widehat{D}^{-1}_{1,n}({\pi}) [\widehat{D}_{2,n}({\pi},{\pi}) +
\widehat{D}_{3,n}({\pi})\widehat{\M{I}}_n^{-1}(\widehat{\V{\varrho}}_n)\widehat{D}_{3,n}^\intercal({\pi})] 
\widehat{D}^{-\intercal}_{1,n}({\pi}).$$
\end{corollary} 
 
 \subsection{Inference for the value of an optimal treatment regime}
 The preceding results establish consistency and asymptotic 
 normality jointly over any fixed set of regimes under the
 estimated MDP.  We now illustrate how these results can be
 used to construct a confidence interval for the value of the
 optimal regime within a (possibly infinite) class of 
 regimes.  We present only the discounted utility
 case as the approach for the average utility is 
 essentially the same.  The strategy we follow here is in the 
 same spirit as in the construction projection confidence intervals,
 which are commonly used   
 for non-smooth functionals
 \citep[see][]{berger1994p,robins2004optimal,laber2014dynamic}.  
 Let $\eta \in (0,1)$ be arbitrary. An overview 
 of the basic approach
  is as follows: (S1) specify 
 a parametric class of regimes; (S2) construct a
 $(1-\eta)\times 100\%$ confidence region for the parameters indexing the 
 optimal regime; mapping each element in this region
 to its corresponding regime thus defines a confidence
 region in the space of regimes; (S3) for each regime
 in the confidence region, construct a $(1-\eta)\times 100\%$ 
confidence interval
 for its value using the asymptotic normality 
 of the estimated value for a fixed regime (derived in the
 previous section); and (S4) take
 a union of all the intervals in the preceding step.  
 It is easily shown that if the region constructed
 in (S2) is a valid confidence region and each interval
 in (S3) is also  (marginally) valid  then the union is a
 valid  $(1-2\eta)\times 100\%$ interval for the optimal value.
For additional discussion see \citet{butchBookDTR}.

Our goal is to derive a confidence interval 
for the optimal value, $\sup_{\pi\in\Pi}V^R_{\mathrm{dis}}({\pi})$,
when ${\Pi}$ is a parametric class of regimes.  We make
the following assumptions. 
 \noindent
\begin{enumerate}
\item [(C9)] The class of regimes ${\Pi}=\{{\pi}(\cdot; \V{\xi}): \V{\xi}\in\M{\Xi}\}$ is indexed by $\V{\xi}$,
where $\M{\Xi}$ is a compact subset of $\mathbb{R}^{\mathrm{dim}\,\V{\xi}}$.
\item [(C10)] The map $\V{\xi} \mapsto V^R_{\mathrm{dis}}(\V{\xi})\triangleq\int\phi^\intercal(\V{s})\V{\alpha}^*(\V{\xi})dR(\V{s})$
has a unique and well-separated maximum at $\V{\xi}=\V{\xi}^*$, which is an interior point of $\M{\Xi}$.
\item [(C11)] There exists a sequence of $\widehat{\V{\xi}}_n\in\M{\Xi}$ such that 
$\widehat{V}_{\mathrm{dis},n}^R(\widehat{\V{\xi}}_n)\geq\sup_{\V{\xi}\in\M{\Xi}}\widehat{V}_{\mathrm{dis},n}^R(\V{\xi}) -o_P(1)$.
\item [(C12)] The map $\V{\xi} \mapsto \int\phi^\intercal(\V{s})\V{\alpha}^*(\V{\xi})dR(\V{s})$ is twice continuously differentiable 
in a neighborhood of $\V{\xi}^*$ and  
$\frac{\partial^2\V{\xi}}{\partial\V{\xi}\partial\V{\xi}^{\T}}
\int \phi^\intercal(\V{s})\V{\alpha}^*(\V{\xi})dR(\V{s})\big |_{\V{\xi}=\V{\xi}^*}$ is positive definite.
\end{enumerate} 
 
\begin{corollary}\label{optparmdiscount}
Assume (A1), (C1) - (C12).  
Then as $n\rightarrow \infty$: 
\begin{enumerate}
\item the results in Theorem \ref{theoremdis2} hold
over all ${\pi} \in {\Pi}$;
\item $\sqrt{n}(\widehat{\V{\xi}}_{n} - \V{\xi}^*)\rightsquigarrow\mathcal{N}(\V{0}, \M{\Sigma}_{\V{\xi}})$,
where $\M{\Sigma}_{\V{\xi}}=\M{\Sigma}_{1,\V{\xi}}^{-1}\M{\Sigma}_{2,\V{\xi}}\M{\Sigma}_{1,\V{\xi}}^{-1}$,
$$\M{\Sigma}_{1,\V{\xi}}= \int\frac{\partial^2}{\partial\V{\xi}\partial\V{\xi}^{\T}}\phi^\intercal(\V{s})\V{\alpha}^*(\V{\xi})dR(\V{s}) \big |_{\V{\xi}=\V{\xi}^*} ,$$ 
$$\M{\Sigma}_{2,\V{\xi}}=\int\biggl\{ \frac{\partial}{\partial\V{\xi}}\phi^\intercal(\V{s})\V{\alpha}^*(\V{\xi}) \biggr\}\biggl\{\frac{\partial}{\partial\V{\xi}} \phi^\intercal(\V{s})\V{\alpha}^*(\V{\xi}) \biggr\}^\intercal dR(\V{s}) 
\big |_{\V{\xi}=\V{\xi}^*}.$$
\end{enumerate}
\end{corollary}

\begin{corollary}\label{projcidiscount}
Assume (A1), (C1) - (C12). Let 
$\eta \in (0, 1)$ be arbitrary.  Define
$$\mathcal{E}_{1-\eta, n}\triangleq \{\V{\xi}: n(\widehat{\V{\xi}}_n-\V{\xi})^\intercal \widehat{\M{\Sigma}}_{\V{\xi},n} (\widehat{\V{\xi}}_n-\V{\xi}) \leq \chi^2_{\mathrm{dim}\,\V{\xi},1-\eta}\},$$
where $\widehat{\M{\Sigma}}_{\V{\xi},n}$ is the sample analog of $\M{\Sigma}_{\V{\xi}}$.
Let $z_{\eta/2}(\V{\xi})$ and $z_{1-\eta/2}(\V{\xi})$ be the $(\eta/2)\times100$ and $(1-\eta/2)\times100$ 
percentiles of a Gaussian distribution with mean-zero and variance $\widehat{\sigma}^{2}_{\mathrm{dis},n}(\V{\xi})$.
Then it follows that 
 $$P \biggl[  \inf_{\V{\xi}\in\mathcal{E}_{1-\eta,n}} \biggl\{ \frac{z_{\eta/2}(\V{\xi})}{\sqrt{n}}  + 
 \widehat{V}_{\mathrm{dis},n}^R(\V{\xi}) \biggr\} \leq V^R_{\mathrm{dis}}(\V{\xi}^*) \leq 
  \sup_ {\V{\xi}\in\mathcal{E}_{1-\eta,n}} \biggl\{ \frac{z_{1-\eta/2}(\V{\xi})}{\sqrt{n}}+\widehat{V}_{\mathrm{dis},n}^R(\V{\xi}) \biggr\} \biggr]\geq 1 - 2\eta.$$ 
\end{corollary}

While projection intervals can be extremely conservative
in some settings \cite[see][]{laber2014dynamic}, in
our simulation experiments, which are based on the STEP-BD study data,  
the degree of conservatism 
is relatively mild. Thus, these intervals appear to be
suitable for application in the settings we consider here.

\section{Simulation experiments}
We study the finite sample performance of the proposed point and interval estimators using 
a series of simulation experiments. The data-generating models we consider
are designed to mimic salient
features of the standard care pathway of the STEP-BD trial.  We consider a follow-up period
of one year.  At each visit, $j\ge 1$, we observe three patient
covariates, $\M{X}^j\in\mathbb{R}^3$, and a treatment is chosen from
among three candidates $A^j\in\lbrace 1, 2, 3\rbrace$ 
so that 
\begin{align*}
\textrm{logit}\biggl\{ \frac{P(A^j=1|\M{H}^j)}{P(A^j=3|\M{H}^j)}\biggr\} &= -0.2 + 0.1X^{j}_1 - 0.1X^{j}_2 + 0.1X^j_3 ,\\
\textrm{logit}\biggl\{ \frac{P(A^j=2|\M{H}^j)}{P(A^j=3|\M{H}^j)}\biggr\} &= -0.2 - 0.1X^{j}_1 + 0.1X^{j}_2 - 0.1X^j_3.
\end{align*}
We consider five latent states intended to encode the health states: depression, mania, mixed type, hypomania,
and stable; thus, $\M{B}^j$ is an element of the five-dimensional probability simplex. 
The $j^{th}$ interarrival time between visits follows an exponential distribution with 
rate $\exp\{e_1 + 0.1(B^j_1+B^j_2) - 0.1(B^j_3+B^j_4)\}$, where $e_1\overset{i.i.d.}{\sim}\textrm{Uniform}(-3,-2)$
is a subject-specific random effect. 
We assume  that the conditional distribution of $\M{X}^{j}$ 
given $(M^j, \M{X}^{j-1})$ follows a Gaussian autoregressive model (see Section 4 
for the form of the density) that is indexed by the following parameters:
\begin{equation*}
\begin{array}{llll}
 \textrm{State } 1: & 
 \V{\mu}_1 = (2,2,2)^{\intercal} & 
\M{\Psi}_1 =  \frac{1}{10}\M{I}_{3\times3} & 
\M{\Sigma}_1= \frac{1}{10}\M{I}_{3\times3}+\frac{1}{10}\M{1}_{3\times3},\\
 \textrm{State } 2: &
\V{\mu}_2 = (2,1,-2)^{\intercal} & 
\M{\Psi}_2 =  \frac{1}{10}\M{I}_{3\times3} &
\M{\Sigma}_2= \frac{1}{10}\M{I}_{3\times3}+\frac{1}{10}\M{1}_{3\times3},\\
 \textrm{State } 3: & 
 \V{\mu}_3 = (-2,1,2)^{\intercal} & 
\M{\Psi}_3 =  -\frac{1}{10}\M{I}_{3\times3} &
\M{\Sigma}_3= \frac{3}{10}\M{I}_{3\times3}-\frac{1}{10}\M{1}_{3\times3},\\
 \textrm{State } 4:& 
\V{\mu}_4 = (-2,-2,-2)^{\intercal} & 
\M{\Psi}_4 =  -\frac{1}{10}\M{I}_{3\times3} &
\M{\Sigma}_4= \frac{3}{10}\M{I}_{3\times3}-\frac{1}{10}\M{1}_{3\times3},\\
 \textrm{State } 5: &
  \V{\mu}_5 = (0,0,0)^{\intercal} &  
\M{\Psi}_5 =  \M{0}_{3\times3} &
\M{\Sigma}_5=  \M{I}_{3\times3},\\
\end{array}
\end{equation*}
where $\V{\mu}_k$, $\M{\Sigma}_k$, and $\M{\Psi}_k$, $k=1,2,\ldots, 5$, are state-dependent mean, covariance,
and autoregression coefficients; $\M{0}_{3\times3}$ is a 3-by-3 matrix of zeros 
and $\M{1}_{3\times3}$ is a 3-by-3 matrix of ones.

In the first scenario, we consider the case where the evolution of latent disease status
follows a first-order Markov process, i.e., the generative model is correctly specified.
We consider the following utility function
\begin{equation*}
U^j = 2 - |X_1^{j+1}| - |X_3^{j+1}|;
\end{equation*}
thus the utility is larger when $X_1$ and $X_3$ are close to 0.
In these simulation experiments, we might think of $X_1$ and $X_3$
as symptom severity measures represented as deviations from 
a stable condition (coded as zero).  
The off-diagonals in the transition rate matrix $\{q_{m,\ell}(a)\}_{m\ne\ell}$ are 
\begin{align*}
&\textrm{logit}\,q_{k,\ell}(a) = e_3 + 5\cdot I(a=1) & \textrm{for }(k,\ell)\in\{(1,5),(4,5),(2,3),(3,2) \},\\
&\textrm{logit}\,q_{k,\ell}(a) = e_3 + 5\cdot I(a=2) & \textrm{for }(k,\ell)\in\{(2,5),(3,5),(1,4),(4,1) \},\\
&\textrm{logit}\,q_{k,\ell}(a) = e_3 + 2\cdot I(a=1) + 2\cdot I(a=2) & \textrm{for }(k,\ell)\in\{(5,1),(5,2),(5,3),(5,4) \},\\
&\textrm{logit}\,q_{k,\ell}(a) = e_3 & \textrm{otherwise,}
\end{align*}
where $e_3\overset{i.i.d}{\sim}\textrm{Uniform}(-7,-6)$ are subject-specific random effects. 
The diagonals are thus $q_{k,k} = -\sum_{\ell\ne k}q_{k,\ell}$ for $k=1,\ldots,5$. 
In this setup, 
treatment 1 will: (1) increase the probability of transitioning to state 5 when the current state is either 1 or 4;
(2) increase the probability of transitioning between state 2 and 3;
and (3) increase the probability of transitioning out of state 5.

In the second scenario, we consider the case where the generative model is misspecified.
At visit $j$, the latent disease states are distributed according to a multinomial distribution
with parameters $(p_1,p_2,p_3,p_4,p_5)$ which are drawn from a Dirichlet distribution with parameter $(1,1,1,1,1)$;
thus, the latent disease state distribution is randomly drawn at each visit.
We consider a utility function of the form
\begin{equation*}
U^j = (B^j_1+B^j_4) \{2 I(A^j=1)-1\} +(B^j_2+B^j_3) \{2 I(A^j=2)-1\} + B^j_5 \{2 I(A^j=3)-1\},
\end{equation*}
which indicates: (1) treatment 1 is optimal when the current state is either 1 or 4;
(2) treatment 2 is optimal when the current state is either 2 or 3;
and (3) treatment 3 is optimal when the current state is 5.
Because this utility is not directly observed, the estimated optimal regime is constructed
with the {\em estimated} utility; however, evaluations are made and reported for the 
true utility. 

We evaluate the mean and standard error of the value 
 of candidate regimes under sample sizes 100 and 200. 
We consider both stochastic and deterministic regimes (stochastic regimes are of interest 
in applications such as mHealth); when estimating
stochastic regimes we used an $L_2$ penalty tuned to ensure that
each treatment is selected with (estimated) probability at least 0.05 across all
observed states \citep[see][]{murphy2016batch}.   
The stochastic regimes we consider include
the data-generating regime, the proposed POMDP regimes in the form of 
multinomial logistic regression using both linear and quadratic basis functions, 
and their MDP regime counterparts, which do not utilize latent state information.
The deterministic regimes we consider include the optimal regime,
the proposed POMDP regimes are linear and indexed by linear and 
quadratic basis functions, and their MDP regime counterparts, which
do not use latent state information.  All results are based on 
500  Monte Carlo replications.

Table \ref{sim1}  shows the mean and standard error for the estimated values 
in scenario 1, where the generative model is correctly specified. 
The proposed POMDP estimators outperform the baseline MDP estimators across
all configurations of stochastic and deterministic regimes and
average and discounted utilities. 
The POMDP estimators have higher mean values and smaller standard
errors than their MDP counterparts. 
Indeed, the values from the estimated POMDP regimes are close to 
those of the true optimal deterministic regime.
In Table \ref{sim2}, where the POMDP model is misspecified,
the estimated values from the POMDP regimes
still significantly outperform the observed and MDP regimes.
This result suggests that
the linear model may be robust to moderate misspecification.  
The inclusion of quadratic terms did not greatly affect performance.
Table \ref{sim3} and Table \ref{sim4} shows the coverage probability and half-width of the proposed confidence
interval for the optimal value under linear regimes when the model is correctly specified
and when the model is misspecified.
The confidence intervals attain nominal (95\%) coverage, although they are a bit conservative as expected.

\begin{table}[H]
\small
\caption{Mean (standard error) for the estimated values for stochastic and deterministic regimes
in scenario 1.}
\centering
\begin{tabular}{lp{1.1cm}p{1.1cm}p{1.1cm}p{1.1cm}p{1.1cm}|p{1.1cm}p{1.1cm}p{1.1cm}p{1.1cm}p{1.1cm}}
\hline
 & \multicolumn{5}{c|}{Stochastic regimes} & \multicolumn{5}{c}{Deterministic regimes} \\
 \cline{2-6}\cline{7-11}
n   &  ${V}_{\mathrm{obs}}$ & ${V}^{\mathrm{MDP}}_{\mathrm{lin}}$ 
&${V}^{\mathrm{MDP}}_{\mathrm{quad}}$ &  
${V}^{\mathrm{POM}}_{\mathrm{lin}}$ & ${V}^{\mathrm{POM}}_{\mathrm{quad}}$  &
${V}^{\mathrm{MDP}}_{\mathrm{lin}}$ &${V}^{\mathrm{MDP}}_{\mathrm{quad}}$ &  
${V}^{\mathrm{POM}}_{\mathrm{lin}}$ & ${V}^{\mathrm{POM}}_{\mathrm{quad}}$ &
 $ V^{\mathrm{opt}} $ \\
\hline
\multicolumn{6}{l|}{Discounted utility}   \\
100 & -4.602 (0.684) & -5.978 (3.517) & -5.888 (3.465) & 1.015 (0.882) & 1.037 (0.851) & -5.878 (3.527) & -5.886 (3.508) & 1.213 (1.060) & 1.158 (1.054) & 1.144 (1.076)    \\
200 & -4.622 (0.527) & -5.321 (3.273) & -5.355 (3.279) & 1.106 (0.584) & 1.062 (0.606) & -5.390 (3.393) & -5.422 (3.359) & 1.265 (0.701) & 1.248 (0.703) & 1.285 (0.689)   \\
\multicolumn{6}{l|}{Average utility}    \\
100 &  -0.481 (0.076) & -0.951 (0.494) & -0.954 (0.500) & 0.032 (0.169) & 0.014 (0.164) & -0.892 (0.448) & -0.893 (0.445) & -0.079 (0.186) & -0.085 (0.181) & -0.063 (0.218)  \\
200 &  -0.483 (0.052) & -0.881 (0.516) & -0.882 (0.518) & 0.053 (0.110) & 0.051 (0.116) & -0.808 (0.444) & -0.807 (0.442) & -0.066 (0.144) & -0.066 (0.135) & -0.017 (0.122) \\
\hline
\end{tabular}
\label{sim1}
\end{table}

\begin{table}[H]
\small
\caption{Mean (standard error) for the estimated values for stochastic and deterministic regimes
in scenario 2.}
\centering
\begin{tabular}{lp{1.1cm}p{1.1cm}p{1.1cm}p{1.1cm}p{1.1cm}|p{1.1cm}p{1.1cm}p{1.1cm}p{1.1cm}p{1.1cm}}
\hline
 & \multicolumn{5}{c|}{Stochastic regimes} & \multicolumn{5}{c}{Deterministic regimes} \\
 \cline{2-6}\cline{7-11}
n   &  ${V}_{\mathrm{obs}}$ & ${V}^{\mathrm{MDP}}_{\mathrm{lin}}$ 
&${V}^{\mathrm{MDP}}_{\mathrm{quad}}$ &  
${V}^{\mathrm{POM}}_{\mathrm{lin}}$ & ${V}^{\mathrm{POM}}_{\mathrm{quad}}$  &
${V}^{\mathrm{MDP}}_{\mathrm{lin}}$ & ${V}^{\mathrm{MDP}}_{\mathrm{quad}}$ &  
${V}^{\mathrm{POM}}_{\mathrm{lin}}$ & ${V}^{\mathrm{POM}}_{\mathrm{quad}}$ & 
$ V^{\mathrm{opt}} $ \\
\hline
\multicolumn{6}{l|}{Discounted utility}   \\
100 & -2.921 (0.168) & 0.412 (0.152) & 0.399 (0.154) & 3.007 (0.153) & 3.009 (0.156) & 0.478 (0.154) & 0.466 (0.157) & 3.660 (0.162) & 3.657 (0.164) & 3.804 (0.091)\\
200 & -2.909 (0.112) & 0.420 (0.107) & 0.420 (0.103) & 3.026 (0.085) & 3.024 (0.083) & 0.482 (0.101) & 0.491 (0.103) &  3.671 (0.076) & 3.672 (0.078) & 3.807 (0.066) \\
\multicolumn{6}{l|}{Average utility}    \\
100 &  -0.343 (0.020) & 0.021 (0.026) & 0.021 (0.024) & 0.262 (0.046) & 0.263 (0.048) & 0.033 (0.025)& 0.031 (0.026) & 0.368 (0.074) & 0.370 (0.074) & 0.418 (0.011) \\
200 & -0.342 (0.014) & 0.024 (0.019) & 0.024 (0.019) & 0.270 (0.036) & 0.271 (0.035) & 0.033 (0.020) & 0.034 (0.021) & 0.384 (0.057) & 0.384 (0.057) & 0.418 (0.008) \\
\hline
\end{tabular}
\label{sim2}
\end{table}

\begin{table}[H]
\small
\caption{Coverage probability and half width of the confidence intervals 
at 0.05 nominal level for the linear POMDP regimes when the model is correctly specified.}
\centering
\begin{tabular}{lccccc}
\hline
& & \multicolumn{2}{c}{Coverage probability} & \multicolumn{2}{c}{Half width} \\
 \cline{3-4}\cline{5-6}
Criterion & n   & Stochastic & Deterministic  & Stochastic  & Deterministic     \\
\hline
 Discounted & 100 & 0.986 & 0.986 & 1.378 & 1.451 \\ 
 Discounted & 200 & 0.988 & 0.984 & 0.966 & 1.010 \\
 \\
 Average & 100 & 0.982 & 0.980 & 0.238 & 0.225 \\
 Average & 200 & 0.978 & 0.990 & 0.175 & 0.185 \\ 
 \hline
\end{tabular}
\label{sim3}
\end{table}

\begin{table}[H]
\small
\caption{Coverage probability and half width of the confidence intervals 
at 0.05 nominal level for the linear POMDP regimes when the model is misspecified.}
\centering
\begin{tabular}{lccccc}
\hline
& & \multicolumn{2}{c}{Coverage probability} & \multicolumn{2}{c}{Half width} \\
 \cline{3-4}\cline{5-6}
Criterion & n   & Stochastic & Deterministic  & Stochastic  & Deterministic     \\
\hline
 Discounted & 100 &  0.978 & 0.972 & 0.315 & 0.289 \\
 Discounted & 200 &  0.974 & 0.970 & 0.192 & 0.181 \\
  \\
 Average & 100 &  0.992 & 0.986 & 0.088 & 0.092 \\
 Average & 200 & 0.978 & 0.982 & 0.071 & 0.062 \\
 \hline
\end{tabular}
\label{sim4}
\end{table}

\section{Case study}

The data used in our case study are derived from the standard care pathway of the STEP-BD clinical trial \citep[][]{sachs2003rationale}.
Inclusion criteria required that patients be (i) at least 18 years old 
and (ii) diagnosed with bipolar type I or bipolar type II disorder at 
screening.
Treatment decisions at each clinic visit were made
based on doctor-patient preference and thus the data
are observational.
Validity of the proposed methods thus requires 
additional causal assumptions. As these assumptions
are standard, we have relegated them to the Supplemental
Material.

\begin{figure}
\begin{adjustbox}{addcode={\begin{minipage}{\width}}{\caption{%
    \label{stepBDTxtHistory}
      Treatment histories for a sample of patients in the 
      STEP-BD standard care pathway. The timing, number, type,
      and dosage of treatment varies widely across patients.  
      }\end{minipage}},rotate=90,center}
      \includegraphics[width=8in]{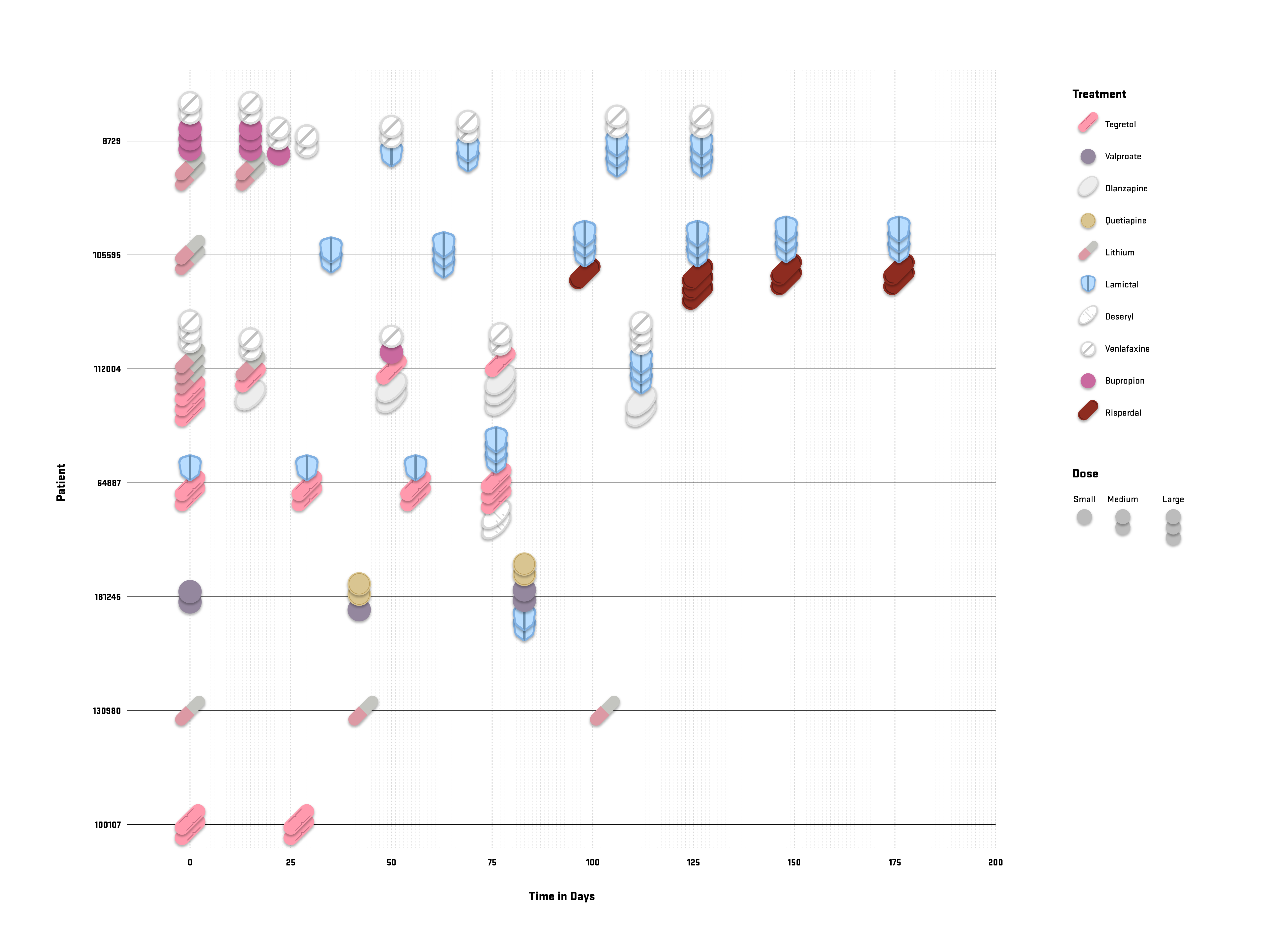}
\end{adjustbox}
\end{figure}

 Figure \ref{stepBDTxtHistory} shows the treatment histories for a sample of patients in the 
 STEP-BD standard care pathway. It can be seen that 
 the timing, number, type,
 and dosage of treatment varies widely across patients. 
We categorize each medication being either an (A) antidepressant or a (M) mood stabilizers; 
and we categorize the dose level for each drug as low, medium, or high. 
The categorization of antidepressants and mood stabilizers as well as
the corresponding dose levels are provided in Appendix I. 
Table \ref{combinations} enumerates the 15 potential treatment combinations.

\begin{table}[H]
\small
\centering
\begin{tabular}{ll}
Treatment ID & Treatment combinations \\
\hline
1 & low A  \\
2 & medium A \\
3 & high A \\
4 & low M \\
5 & medium M \\
6 & high M \\
7 & low A $+$ low M  \\
8 & low A $+$ medium M \\
9 & low A $+$ high M \\
10 & medium A $+$ low M \\
11 & medium A $+$ medium M  \\
12 & medium A $+$ high M \\
13 & high A $+$ low M  \\
14 & high A $+$ medium M \\
15 & high A $+$ high M  \\
 \hline
\end{tabular}
\caption{List of potential treatment combinations. A = antidepressants, M = mood stabilizers.}
\label{combinations}
\end{table}

We assume that there are five latent health states corresponding to: depression, mania, mixed type, hypomania,
and stable moods.
At each stage, a patient's estimated state comprises the
latent health state probability vector and observable patient covariates:
age, bipolar disorder type, 
sum of depression score (SUMD), 
sum of mania score (SUMM), 
percent of days depressed, 
percent of days low interest in most activities, 
and percent of days with abnormal mood elevation. 
SUMD and SUMM are aggregates of multiple items on a questionnaire.
In the study protocol, both SUMD and SUMM are defined to be missing when an answer
to any of the inventory questions is missing, this results in 19\% and 36\% missing entries, respectively.
We used multiple imputation \citep[][]{rubin2004multiple} for missing items
and recalculated aggregated scores using the imputed data.
Besides the inventory questions in SUMD and SUMM, other variables used in multiple imputation
include the three other continuous patient covariates as well as patient baseline
characteristics (details and code are provided in Supplemental Material).
We imputed five complete data sets, and to each imputed
data set we applied the proposed
methodology to estimate the optimal treatment regime. Parameters indexing
each estimated optimal treatment regime 
were averaged and used in the
final estimated optimal treatment regime.  
The utility at each stage is defined as 
$2-\textrm{SUMD}-\textrm{SUMM}$, 
where both SUMD and SUMM are standardized to lie between zero and one.
A higher utility implies a lower SUMD and SUMM, which corresponds to a more desirable clinical outcome.
\cite{wu2015will} used SUMD as the clinical outcome in estimating the optimal treatment regime
in the randomized arm of STEP-BD, where there were only two decision stages. 
However, an effective long-term treatment regime for bipolar disorder should alleviate
depression symptoms without inducing manic episodes, which is why we opted for a composite outcome. 

Table \ref{clinstat} shows the mean and standard error of the estimated mood state probabilities using the proposed 
latent Markov model.  The results are promising in that they 
largely agree with the reported clinical status on the clinical monitoring form.\footnote{Such assessments
were collected in the trial and thus can serve as a kind of gold standard. However, these are 
not collected as a matter of course in standard clinical care which is why they were not used in the modeling. 
In cases where 
such assessments are made at each visit, they can be folded into the observed state as 
noisy surrogate for the true latent state at the visit time.}
The model had some difficulty delineating between mania and hypomania; however, this is not surprising
as (clinically) these abnormal states differ only in severity.  

Table \ref{value} shows the value under the observed regime and under the estimated regime for
average utility and discounted utility ($\gamma=0.95$). 
In both cases, the lower bound of the 95\% confidence interval for the value of the estimated regime
is higher than the observed regime.

\begin{figure}[H]
      \includegraphics[width=7in]{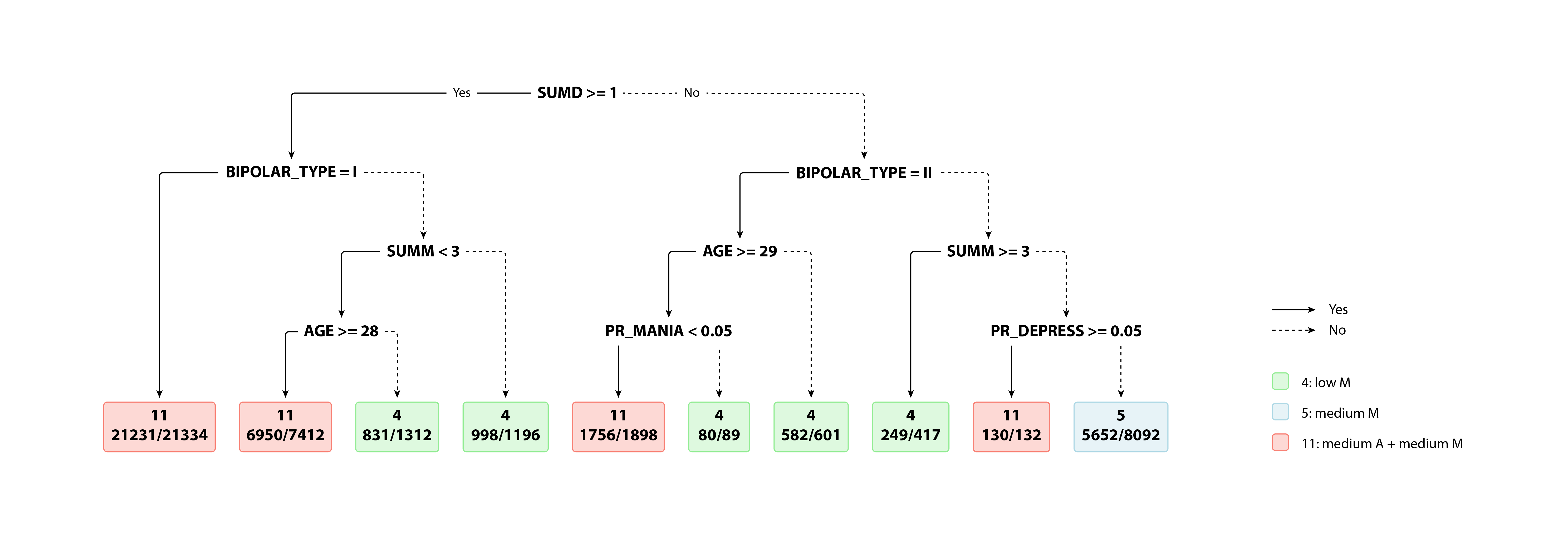}
      \caption{Estimated optimal treatment regime obtained by maximizing average utility, 
    which is projected onto a decision tree for interpretation. Each tree node shows the splitting criterion,
    majority treatment label, and its proportion.} 
  \label{stepbd_tree2}
\end{figure}

Figure \ref{stepbd_tree2}  shows the estimated optimal treatment regime  obtained by maximizing the average utility,
which is projected onto a decision tree for ease of interpretation. 
The predicted optimal treatment is either mood stabilizers 
or a combination of mood stabilizers and antidepressants, i.e., it 
never recommends antidepressants alone.    
Such recommendations are anticipated by the clinical belief that prescribing antidepressants 
alone for bipolar disorder patients 
may increase the risk of inducing a manic episode \citep[][]{patel2015antidepressants}.
The estimated optimal regime also prescribes antidepressants only as a supplement for the mood stabilizer 
either when there is some evidence of depression (SUMD is large or the probability of depression is large)
or there is little evidence of mania (SUMM is small or the probability of mania is small).
The estimated optimal treatment regime for the discounted utility ($\gamma=0.95$) is included in the Appendix and 
is qualitatively similar.

\begin{table}[H]
\small
\centering
\begin{tabular}{llllll}
Clinical status & $\widehat{P}$(Depress) &  $\widehat{P}$(Mania)  & $\widehat{P}$(Mixed)  & $\widehat{P}$(Hypomania) &$\widehat{P}$(Stable) \\
\hline
Depression & 0.85  & $<$0.01  & 0.14 & $<$0.01 & 0.05 \\
Mania &  $<$0.01 & 0.55 &0.05  & 0.47  & $<$0.01\\
Mixed &     0.13  &0.09 & 0.84  & 0.08   & $<$0.01\\
Hypomania & $<$0.01 &0.34 & 0.09  & 0.44  & 0.02\\
Stable & 0.02  &$<$0.01 & $<$0.01 & $<$0.01   & 0.92\\ \hline
\end{tabular}
\caption{Mean estimated mood state probabilities within each of the five clinical status categories.
The standard errors range from 0.0001 to 0.01.}
\label{clinstat}
\end{table}

\begin{table}[H]
\small
\centering
\begin{tabular}{lll}
Criterion & Observed value & Value under estimated regime (95\% C.I.) \\
\hline
 Average utility & 1.66 & 1.81 (1.72, 1.88) \\
 Discounted utility & 14.40 & 33.76 (19.24, 46.95) \\ \hline
\end{tabular}
\caption{Comparison of the value under the observed regime and the value under the estimated regime
for average utility and discounted utility ($\gamma=0.95$).}
\label{value}
\end{table}

\section{Conclusions}

We developed a  framework
for estimation of an optimal treatment regime using data from 
long-term observational or randomized clinical studies. From
a precision medicine perspective, a
key contribution of this work is the incorporation of a patient's 
latent health status, e.g., their true mood state in the context
of bipolar depression.  We showed that using this structure 
can lead to estimated optimal regimes that are clinically meaningful
and that significantly outperform methods that fail to use this structure.
From a methodological perspective, a key contribution is the development of 
methods for estimation and 
inference for the optimal treatment regime using from data
generated from a POMDP.    
One could generalize the proposed methodology to include
continuous latent processes. Such an approach is aligned with existing work 
in POMDPs in the computer science and engineering literature. 
We leave such extensions
to future work.

 \section*{Appendix I: Tables for medications}

 \begin{table}[H]
\small
\centering
\begin{tabular}{llll}
Medication name & Low dose (mg) & Medium dose (mg) & High dose (mg)\\
\hline
Deseryl  &  $<200$ & $200-400$ & $>400$ \\
Serzone  & $<200$ & $200-400$ & $>400$\\
Citalopram &   $<20$ & $20-40$ & $>40$\\
Escitalopram Oxalate &  $<10$ & $10-20$ &$>20$\\
Prozac  & $<20$ &  $20-40$ & $>40$\\
Fluvoxamine  & $<100$ & $100-200$ & $>200$\\
Paroxetine  & $<20$ & $20-40$ & $>40$\\
Zoloft &  $<50$ & $50-100$ & $>100$\\
Venlafaxine &  $<75$ & $75-150$ & $>150$\\
Bupropion &  $<150$ & $150-300$ & $>300$\\
\hline
\end{tabular} 
\caption{List of common antidepressants in STEP-BD. The 
dose is divided into 3 levels: low, medium, and high.}
\label{antidepressants}
\end{table}

\begin{table}[H]
\small
\centering
\begin{tabular}{llll}
Medication name & Low dose (mg) & Medium dose (mg) & High dose (mg) \\
\hline
 Tegretol &   $<400$ & $400-800$ & $>800$\\
 Valproate &   $<1000$ & $1000-2000$ & $>2000$\\
  Olanzapine  & $<10$ & $10-20$ & $>20$\\
Quetiapine &  $<400$ &  $400-800$ & $>800$\\
Clozapine &  $<200$ &  $200-400$ &  $>400$\\
Lithium &  $<900$ & $900-1800$ &$>1800$\\
Risperdal & $<2$ & $2-4$ &  $>4$\\
Geodon &  $<80$ &  $80-160$ &  $>160$\\
Abilify & $<15$ & $15-30$ &  $>30$\\
Lamictal &   $<100$ & $100-200$ & $>200$\\
\hline
\end{tabular}
\caption{List of common mood stabilizers in STEP-BD. The 
dose is divided into 3 levels: low, medium, and high.}
\label{moodstabilizers}
\end{table}

\section*{Appendix II: Estimated optimal treatment regime for total discounted utility}

\begin{figure}[H]
      \includegraphics[width=7in]{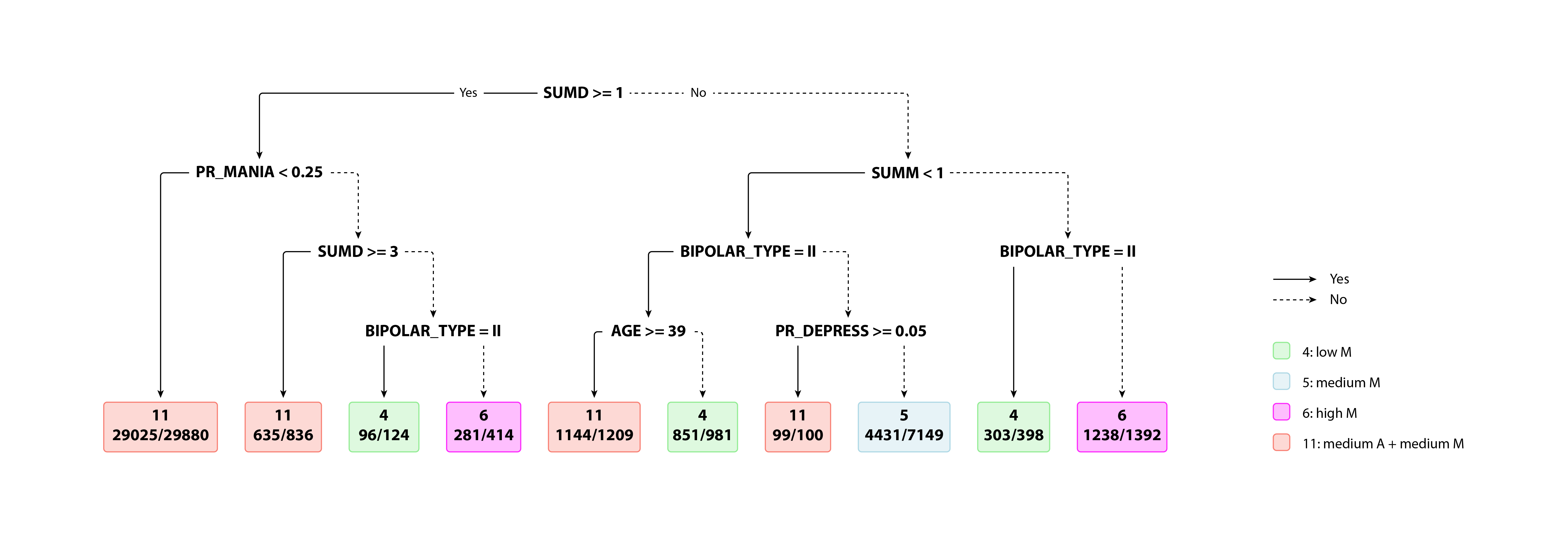}
      \caption{Estimate optimal treatment regime by maximizing total discounted utility ($\gamma=0.95$), 
    which is projected onto a decision tree for interpretation. Each tree node shows the splitting criterion,
    majority treatment label and its proportion.} 
    \label{stepbd_tree1}
\end{figure}

\bibliographystyle{Chicago}

\bibliography{bibliograph}

\end{document}